\documentclass[journal,lettersize]{IEEEtran}

\pdfpageattr{/Group << /S /Transparency /I true /CS /DeviceRGB >>}

\usepackage[utf8]{inputenc}
\usepackage[T1]{fontenc}
\usepackage{textcomp}
\usepackage{stfloats}
\usepackage{fixltx2e}
\usepackage{microtype}
\usepackage{calc}
\usepackage[normalem]{ulem}

\usepackage[range-phrase=--,per-mode=symbol-or-fraction,binary-units=true,range-units=single,list-units=single,detect-all,forbid-literal-units]{siunitx}
\usepackage{silence}
\AtBeginDocument{
\DeclareSIUnit\kmh{\km\per\hour}
\DeclareSIUnit\mps{\m\per\s}
\DeclareSIUnit\vehicle{veh}
\DeclareSIUnit\lane{lane}
\DeclareSIUnit\flow{\vehicle\per\hour}
\DeclareSIUnit[per-mode=repeated-symbol]{\density}{\vehicle\per\km\per\lane}
\newcommand{\densityCaption}{\thinspace{}veh/km/lane} 
\DeclareSIUnit[per-mode=repeated-symbol]{\lanecapacity}{\vehicle\per\lane\per\hour}
\DeclareSIUnit\ghz{\giga\hertz}
\DeclareSIUnit\hz{\hertz}
\DeclareSIUnit\km{\kilo\metre}
\DeclareSIUnit\m{\metre}
\DeclareSIUnit\s{\second}
}

\usepackage{csquotes}
\usepackage[backend=bibtex,style=ieee,citestyle=numeric-comp,doi=false,isbn=false,mincitenames=1,maxcitenames=2,minbibnames=1,maxbibnames=5]{biblatex}
\DeclareFieldFormat{sentencecase}{#1} 
\DeclareFieldFormat{titlecase}{#1} 
\usepackage{xpatch}
\xpatchbibmacro{Textcite}{\addspace}{\addnbspace}{}{}
\DefineBibliographyStrings{english}{
    andothers = et~al\adddot\addspace
}

\usepackage{amsmath}

\usepackage{amssymb}
\usepackage{amsfonts}
\usepackage{amsthm}

\usepackage{booktabs}
\usepackage{url}
\usepackage{multirow}

\usepackage[inline]{enumitem}

\usepackage[caption=false,font=footnotesize]{subfig}
\usepackage[pdftex]{graphicx}
\usepackage{pgfplots}
\DeclareGraphicsExtensions{.pdf,.png,.jpg}

\usepackage{todonotes}

\usepackage{tikz}
\usetikzlibrary{arrows}
\usetikzlibrary{calc}
\usetikzlibrary{chains}
\usetikzlibrary{scopes}

\usepackage[american]{babel}
\usepackage{hyphenat}

\usepackage[noend]{algorithmic}
\usepackage{algorithm}

\usepackage{hyperref}
\hypersetup{pdftex,colorlinks=true,allcolors=blue}
\usepackage{hypcap}

\usepackage[capitalize,noabbrev]{cleveref}
\crefname{paragraph}{Paragraph}{Paragraphs}
\Crefname{paragraph}{Paragraph}{Paragraphs}

\usepackage[nodebug]{flushend}

\RequirePackage{xstring}
\RequirePackage{xparse}
\RequirePackage[]{acro}
\NewDocumentCommand\acrodef{mO{#1}mG{}}{\DeclareAcronym{#1}{short={#2}, long={#3}, #4}}
\acrodef{ACC}{Adaptive Cruise Control}
\acrodef{ADAS}{Advanced Driver Assistance System}
\acrodef{AHS}{Advanced Highway System}
\acrodef{CACC}{Cooperative Adaptive Cruise Control}{alt={Cooperative Adaptive Cruise Controller}}
\acrodef{CAM}{Cooperative Awareness Message}
\acrodef{CC}{Cruise Control}{alt={Cruise Controller}}
\acrodef{CF}{Car Following}
\acrodef{C-V2X}{Cellular \acl*{V2X}}
\acrodef{DG}{Destination Group}
\acrodef{DGPS}{Dynamic Grouping and Platoon Splitting}
\acrodef{DSRC}{Distributed Short-Range Communication}
\acrodef{DYG}{Dynamic Grouping}
\acrodef{eCDF}{empirical Cumulative Distribution Function}
\acrodef{ETSI}{European Telecommunications Standards Institute}
\acrodef{HBEFA}{Handbook Emission Factors for Road Transport}
\acrodef{IDM}{Intelligent Driver Model}
\acrodef{ITS}{Intelligent Transportation System}{short-plural-form={ITS}}
\acrodef{IVC}{Inter-Vehicle Communication}
\acrodef{MIP}{Mixed Integer Programming}
\acrodef{PCS}{Platooning Coordination System}
\acrodef{RA}{Random Assignment}
\acrodef{ROI}{Region of Interest}{short-indefinite={an}, long-plural-form={Regions of Interest}}
\acrodef{RSU}{Roadside Unit}{short-indefinite={an}}
\acrodef{TPFS}{Transient Platoon Formation Strategy}
\acrodef{V2I}{Vehicle-to-Infrastructure}
\acrodef{V2V}{Vehicle-to-Vehicle}
\acrodef{V2X}{Vehicle-to-Everything}
\acrodef{VANET}{Vehicular Ad Hoc Network}
\acrodef{VENTOS}{VEhicular NeTwork Open Simulator}
\acrodef{VLC}{Visible Light Communication}
\acrodef{V-VLC}{Vehicular \acl*{VLC}}

\usepackage[color]{changebar}
\def\todo{%
    \cbcolor{magenta}
    \cbstart%
    \begingroup
    \color{magenta}
    \obeylines%
    \begingroup\lccode`~=`\^^M\lowercase{\endgroup\def~}{\par\leavevmode}%
    \parindent0em%
    \catcode`\_=\active
    \catcode`\<=\active\lccode`~=`<\lowercase{\def~}{$<$}%
    \catcode`\>=\active\lccode`~=`>\lowercase{\def~}{$>$}%
    \catcode`\#=\active\lccode`~=`\#\lowercase{\def~}{$\#$}%
    \catcode`\^=\active\lccode`~=`\^\lowercase{\def~}{$\hat{~}$}%
    \todoCtd
}\def\todoCtd#1{%
    TODO: #1%
    \ifx&#1&...\fi%
    \endgroup
    \cbend
    \relax
}
\usepackage{comment}
\usepackage{draftfigure}

\NewDocumentCommand\IEEE{ s m d[] }{%
    \IfBooleanTF{#1}{}{IEEE\,}
    \nolinebreak[2]
    #2%
    \IfNoValueTF{#3}{%
    }{%
        \StrGobbleLeft{#3}{1}[\sommerIEEEFirstLetter]%
        \IfEq{\sommerIEEEFirstLetter}{}{%
            #3
        }{%
            \nolinebreak[3]
            \StrLeft{#3}{1}%
            \sommerIEEELettersSlashed{\sommerIEEEFirstLetter}%
        }%
    }%
}
\newcommand{\sommerIEEELettersSlashed}[1]{%
    /
    \StrLeft{#1}{1}%
    \StrGobbleLeft{#1}{1}[\sommerIEEESubsequentLetter]%
    \IfEq{\sommerIEEESubsequentLetter}{}{%
    }{%
        \sommerIEEELettersSlashed{\sommerIEEESubsequentLetter}
    }%
}

\newcommand{\plexe}{Plexe}
\newcommand{\sumo}{SUMO}
\newcommand{\simulator}{PlaFoSim}
\newcommand{\p}{\IEEE{802.11}[p]}

\newcommand{\vtp}{vehicle-to-platoon}
\newcommand{\human}{\emph{human}}
\newcommand{\acc}{\emph{\acs*{ACC}}}
\newcommand{\distributed}{\emph{distributed greedy}}
\newcommand{\centralized}{\emph{centralized greedy}}
\newcommand{\optimal}{\emph{centralized solver}}

\newcommand{\fc}{\ensuremath\tilde{g}}

\begin{document}

\title{%
%
%
%
Where to Decide? Centralized vs.\ Distributed\\Vehicle Assignment for Platoon Formation
}

\author{%
\IEEEauthorblockN{%
    Julian Heinovski\IEEEmembership{~Graduate~Student~Member,~IEEE}%
    ~%
    and
    Falko Dressler\IEEEmembership{~Fellow,~IEEE}%
}%
\thanks{%
Julian Heinovski and Falko Dressler are with the School of Electrical Engineering and Computer Science, TU Berlin, Germany;
E-Mails: \{heinovski,dressler\}@ccs-labs.org.
}%
}%

\maketitle

\begin{abstract}\nohyphens{%
Platooning is a promising cooperative driving application for future intelligent transportation systems.
In order to assign vehicles to platoons, some algorithm for  platoon formation is required.
Such \vtp{} assignments have to be computed on-demand, e.g., when vehicles join or leave the freeways.
In order to get best results from platooning, individual properties of involved vehicles have to be considered during the assignment computation.
In this paper, we explore the computation of \vtp{} assignments as an optimization problem based on similarity between vehicles.
We define the similarity and, vice versa, the deviation among vehicles based on the desired driving speed of vehicles and their position on the road.
We create three approaches to solve this assignment problem: \optimal{}, \centralized{}, and \distributed{}, using a \ac{MIP} solver and greedy heuristics, respectively.
Conceptually, the approaches differ in both knowledge about vehicles as well as methodology.
We perform a large-scale simulation study using \simulator{} to compare all approaches.
While the \distributed{} approach seems to have disadvantages due to the limited local knowledge, it performs as good as the \optimal{} approach across most metrics.
Both outperform the \centralized{} approach, which suffers from synchronization and greedy selection effects.
The \optimal{} approach however assumes global knowledge and requires a complex \ac{MIP} solver to compute \vtp{} assignments.
Overall, the \distributed{} approach achieves close to optimal results but requires the least assumptions and complexity.
Therefore, we consider the \distributed{} approach the best approach among all presented approaches.
}\end{abstract}


\acresetall%
\IEEEpeerreviewmaketitle%

%

\section{Introduction}%
\label{sec:introduction}

\IEEEPARstart{T}{he} amount of road traffic has been constantly growing in recent years, leading to more and more congestion of the roads and environmental pollution.
To cope with these negative effects, vehicle manufacturers and researchers are striving to improve today's driving to be more efficient and comfortable.
Vehicles are equipped with more and more technology to assist the driver in his tasks in order to make driving more efficient and safe, transforming them into \acp{ITS}.
Using \ac{IVC} technologies like 5G-based \ac{C-V2X} communication or \ac{DSRC}, vehicles are now able to cooperate with each other.
This allows applications such as cooperative driving~\cite{dressler2019cooperative,locigno2022cooperative}.

One such application is vehicular platooning where multiple vehicles are grouped into a convoy and electronically coupled via \ac{CACC}
\cite{%
,shladover1991automated%
,bergenhem2012overview%
,jia2016survey%
}.
This coupling allows small safety gaps (e.g., \SI{5}{\metre}) between platoon members while still enabling string-stable and safe operation.
Platooning thus promises to enhance today's driving by improving traffic flow~\cite{vanarem2006impact}.
%
The technical feasibility of string-stable platooning has been investigated in depth in the literature
\cite{%
,shladover1991automated%
,horowitz2000control%
,robinson2010operating%
,jootel2012final%
,jia2016survey%
}.
Most studies typically consider pre-configured and well-defined platoons.
This assumption, however, is somewhat unrealistic as some form of bootstrapping a platoon is required:
Vehicles will initially drive individually until they encounter an appropriate (existing) platoon to join or another individual vehicle to form a new platoon with.
In order to perform the necessary driving tasks (e.g., approaching, lane changing, switching to \ac{CACC}) to become a platoon member, cooperative maneuvers need to be executed
\cite{%
,lam2013cooperative%
,segata2014supporting%
}.
Deciding which vehicles should form a platoon together is part of the general challenge of assigning vehicles to platoons, which requires computing so called \emph{\vtp{} assignments}.

While simple ad-hoc approaches enable fast setup of platooning~\cite{maiti2019analysis}, \vtp{} assignments typically require more complex computation in order to optimize for certain factors~\cite{sturm2020taxonomy}.
Some first solutions for computing such assignments in an optimal way are limited to, e.g., a certain set of vehicles with known trips and requirements, allowing a priori computations~\cite{bhoopalam2018planning}.
However, the trips and requirements of vehicles are not always known, which requires on-demand and en route computation of assignments based on the current situation.
Thus, new vehicles can only be considered for platooning after entering the freeway and announcing their interest in platooning.
Only now, vehicles or other entities can start searching for appropriate platoon candidates and compute corresponding assignments.
While there are ideas to wait for other vehicles to form platoons before entering the actual freeway~\cite{hall2005sorting}, doing so will unnecessarily delay the trips.
Therefore, the entire platoon formation process including the computation of \vtp{} assignments should happen during the vehicles' trips on the freeway.
%
To avoid assigning vehicles to platoons with heterogeneous properties, the individual vehicles properties should be considered during the assignment computation~\cite{lesch2021overview,heinovski2018platoon}.
Thus, the platooning benefits for the individual vehicles can be optimized, without giving up too much of the own requirements, instead of system-level aspects such as traffic flow.
Therefore, solving the challenge of \vtp{} assignments with respect to these aspects is the next important step towards large-scale deployment of platooning~\cite{hou2023large-scale}.

In this paper, inspired by our earlier work~\cite{heinovski2018platoon}, we explore the computation of \vtp{} assignments as an optimization problem based on similarity between vehicles.
We define the similarity and, vice versa, the deviation among vehicles based on the desired driving speed of vehicles and their position on the road, thereby considering their individual requirements.
We aim to increase vehicles' similarity, thereby minimizing their deviation in desired driving speed and position.
We create three approaches to solve this assignment problem: \optimal{}, \centralized{}, and \distributed{}, using a \ac{MIP} solver and greedy heuristics, respectively.
Conceptually, the approaches differ in both knowledge about vehicles as well as methodology.
We assume that such coordination only requires limited network access, which can easily be provided by modern technologies such as 5G-based \ac{C-V2X} or \p{}-based \ac{DSRC}.
We perform a large-scale simulation study with \simulator{} to compare both greedy approaches to the optimal solution.
We report extensive simulation results to evaluate the impact of the knowledge as well as the performance of the approximation by the heuristics.

Our main contributions can be summarized as follows:
\begin{itemize}
    \item We explore the problem of \vtp{} assignments as an optimization problem based on similarity between vehicles.
    \item We developed three approaches to solve this optimization problem: \optimal{}, \centralized{}, and \distributed{}, using a \ac{MIP} solver and greedy heuristics, respectively.
    \item We perform a large-scale simulation study with \simulator{} to compare both greedy approaches to the optimal solution.
    \item We show that the \distributed{} approach leads to the best results while requiring the least assumptions and complexity among all presented approaches.
\end{itemize}

The remainder of this paper is structured as follows:
First, we review related work from the literature in \cref{sec:related}.
We then propose our idea of computing \vtp{} assignments based on similarity between vehicles in \cref{sec:platoon_formation}.
Afterwards, we illustrate our methodology for evaluating the proposed idea and report corresponding results in \cref{sec:eval}.
We discuss and summarize the results of our evaluation in \cref{sec:discussion}.
Finally, we summarize and conclude our findings in \cref{sec:conclusion}.

%

\section{Related Work}%
\label{sec:related}

\noindent
A simple approach of forming platoons is to perform spontaneous (ad-hoc) platoon formation with other vehicles, using only limited consideration of vehicles' properties and avoiding complex decision making~\cite{maiti2019analysis}.
Vehicles can join close platoons based on their current position without considering a particular constraint or properties of other vehicles~\cite{maiti2019analysis}.
Adding a little more complexity, vehicles can evaluate whether the estimated benefit of catching up or slowing down and joining a platoon is more fuel-efficient than driving alone~\cite{liang2013when}.
Combining catch-up and slow-down, \textcite{saeednia2016analysis} propose a hybrid platooning strategy targeting the formation of truck platoons with the highest possible platooning speed.
\textcite{woo2021flow-aware} propose a flow-aware strategy for platoon organization, which performs formation conditionally on the local traffic state (i.e., flow and speed) in order to avoid degrading its performance.

Ad-hoc approaches often form platoons under the assumption that platooning is desired and to improve macroscopic metrics such as lane capacity and traffic flow.
However, when considering a more microscopic level, i.e., metrics that directly influence and are of interest to a driver, this might not be optimal as individual properties and capabilities of the trips and vehicles, i.e., destination, desired driving speed, trip duration, fuel consumption, are not considered.
In the following, we separately report on centralized, decentralized, and distributed platoon formation solutions in the literature.

\subsection{Centralized Coordination}%
\label{sec:rw_centralized}

\noindent
In \emph{centralized} coordination, \vtp{} assignments are computed from a global perspective for all vehicles.
Thus, it is assumed that details such as trip information are known to the coordination system.

Many studies consider platooning of trucks, where trips and schedules or deadlines are known beforehand.
This allows to use static planning models, where a central coordination instance computes \vtp{} assignments based on trucks' travel information a priori~\cite{bhoopalam2018planning}.
Considering trucks from a single fleet, many studies propose approaches for offline optimization of the trucks' fuel consumption by using platooning
\cite{%
,larsson2015vehicle,%
,vandehoef2015fuel-optimal%
,vandehoef2015coordinating%
,nourmohammadzadeh2018fuel%
}.
In addition to the \vtp{} assignments, the algorithms compute speed profiles, routes, and departure times~\cite{sokolov2017maximization} for the trucks in order to fulfill the desired platoons.
Recently, such optimization was proposed also for trucks from multiple fleets with different objectives, utilizing the trucks' computational resources~\cite{zeng2022decentralized}.
The complexity of such centralized optimization has been shown to be NP-hard~\cite{larsson2015vehicle}.

Since these offline approaches are limited to known sets of vehicles including their properties and trip data, they do not work well for general traffic.
Thus, on-demand and en route computation of \vtp{} assignments based on the current situation on the freeway is required.
%
\textcite{liang2016heavy-duty} study fuel-efficient platooning for trucks and form platoons on the fly.
They consider an optimization problem for pairwise coordination of vehicles in a centralized manner, where the leading vehicle slows down and the trailing vehicle speeds up.
\textcite{vandehoef2018fuel-efficient} define a combinatorial optimization problem that uses the transport assignments of trucks for en route formation of truck platoons.
The process is repeated whenever assignments change or deviation from the plans are detected.

Considering normal passenger vehicles, \textcite{krupitzer2017rocosys} propose a centralized platoon coordination system that receives information (i.e., destination) from drivers via \ac{V2X} communication, searches for a feasible platoon, and performs the corresponding join maneuver.
\textcite{liu2022optimizing} split the highway into zones of \SI{2}{\km} and propose an optimization problem that decides whether or not each single vehicle should join a specific platoon within each zone.

%

\subsection{Decentralized Coordination}%
\label{sec:rw_decentralized}

\noindent
In \emph{decentralized} coordination, \vtp{} assignments are computed in multiple locations (e.g., \aclp{RSU}) from limited global perspectives for a subset of vehicles.
Thus, decentralized approaches have only limited global knowledge about an area (e.g., a road section) or other vehicles in \ac{V2X} communication proximity.
%
Several studies used a decentralized approach that sorts vehicles into platoons at the entrance ramp of the freeway~\cite{hall2005sorting,baskar2008dynamic}.
These approaches group vehicles according to no or limited constraints (e.g., their destination) and let them enter the freeway as an already constructed platoon at a certain time.
%
Like ramp-grouping, platoons can be formed with trucks waiting at hubs along the freeway.
\textcite{larsen2019hub-based,johansson2020truck} propose to use local coordinators at hubs for dispatching multiple trucks at the same time in order to form a platoon.
The goal is to optimize fuel savings and minimize the cost of waiting at the hub, while adhering to the waiting time windows of trucks.


%
\citeauthor{larson2013coordinated}\cite{larson2013coordinated,larson2015distributed} deploy a distributed network of local controllers at junctions in a road network.
The controllers monitor approaching trucks and coordinate platoon formation among all vehicles in proximity.
Using vehicles' information such as speed, position, and destination, an optimization problem for required speed adjustments and expected fuel savings is solved via heuristics.
%
Other studies divide the highway into sections and coordinate platooning within those sections with local centralized controllers.
%
\textcite{krupitzer2018towards} extend their centralized platoon coordination system with sub-systems that individually coordinate vehicles within sections in a regional planner-like approach.
While only limited details are provided, their approach uses \ac{V2I} communication for exchanging vehicles' desired driving speed and route data with the coordination system, which forms platoons among vehicles with a similar route or destination.
Similarly, \textcite{burov2020platoon} use local centralized controllers for highway sections of \SI{300}{\m} to coordinate platooning.
The controllers collect and transmit data between platoons and vehicles via \ac{V2I} communication.
%
Going one step further, \textcite{zhu2022hindrance-aware} use local controllers to provide general driving guidance and platoon coordination for vehicles within a controlling zone (i.e., a freeway section).
They create an optimization problem for jointly considering speed, safety, and energy efficiency, in order to maximize overall traffic velocity and minimize collision risk and fuel consumption.

%

\subsection{Distributed Coordination}%
\label{sec:rw_distributed}

\noindent
In \emph{distributed} coordination, \vtp{} assignments are computed within the individual vehicles.
Here, the vehicle itself is the actor taking the decision about forming a platoon with other vehicles.
Thus, distributed approaches have only limited local knowledge about other vehicles in \ac{V2X} communication proximity.
%
\textcite{khan2005convoy} develop a system which evaluates the cost and benefit of forming a platoon with other vehicles in proximity.
The system continuously evaluates the utility of platooning in terms of deviation from desired driving speed and fuel consumption; if successful, it indicates the decision to the driver using an LED to adjust the \ac{ACC} accordingly.
Since vehicles' communication range is limited to \SI{20}{\m}, vehicles continue evaluation after joining a platoon, allowing them to switch to a platoon with higher utility if encountered.

One way of grouping vehicles is based on their destination, thereby maximizing the distance a platoon stays intact and vehicles can share platoon benefits~\cite{dao2008decentralized,dao2013strategy}.
Vehicles on entrance ramps of a freeway use \ac{V2X} to communicate with other vehicles and platoons in range to find feasible platooning opportunities.
They select a platoon with members that have similar destinations (within a certain range), thereby increasing lane capacity and traffic throughput~\cite{dao2008decentralized} as well as fuel consumption~\cite{dao2013strategy}.
If vehicles cannot immediately find a feasible platoon, they can temporarily join a non feasible one and perform a one-time change to a better fitting platoon later~\cite{hobert2012study}.
Also using vehicles' destination and route, \textcite{dokur2022edge} propose a system that exchanges this data with broadcasts via \ac{DSRC}.
If vehicles are heading to the same destination or to different destinations but share a common path, they start negotiations for forming a platoon via messages.
A negotiation resolver running on the vehicles settles the negotiations between the vehicles to initiate platooning, also deciding about the platoon leader.

In contrast, many studies use vehicles' speed and position for forming platoons.
\textcite{su2016autonomous} propose a distributed algorithm that uses other vehicles' speed and position data from \p{} broadcasts.
Vehicles calculate the difference between their own value and the average of all neighbors in speed and position.
They also choose suitable platoon leaders for new platoons based on an exponential distribution of the speed difference and announce their result using a slotted persistence approach.


%

\subsection{Open Research Problems}%
\label{sec:open_problems}

\setlength{\tabcolsep}{3pt}
\begin{table*}[t]
\centering
\caption{Categorization of related work with on-demand \vtp{} assignment computation.}%
\label{tab:rw}
\begin{tabular}{|l|cc|cc|cc|cc|c|c|}
\hline
\multirow{2}{*}{Related Work} &
\multicolumn{2}{c|}{assignment time} &
\multicolumn{2}{c|}{vehicle knowledge} &
\multicolumn{2}{c|}{coordination} &
\multicolumn{2}{c|}{solution} &
considered \\
                                                        & before trip & en route & global & local & centralized & distributed & optimal & heuristic & properties                    \\
\hline
\textcite{liang2016heavy-duty}                          &             & x        & x      &       & x           &             & x       &           & route, deadline               \\
\textcite{vandehoef2018fuel-efficient}                  &             & x        & x      &       & x           &             & x       &           & route, deadline               \\
\textcite{krupitzer2017rocosys}                         &             & x        & x      &       & x           &             &         &           & destination                   \\
\textcite{liu2022optimizing}                            &             & x        & x      &       & x           &             & x       &           & speed, destination            \\
\textcite{hall2005sorting}                              & x           &          &        & x     & x           &             &         & x         & destination                   \\
\textcite{baskar2008dynamic}                            & x           &          &        & x     & x           &             &         & x         & destination                   \\
\textcite{larsen2019hub-based}                          & x           &          &        & x     & x           &             &         & x         & route, deadline               \\
\textcite{johansson2020truck}                           & x           &          &        & x     & x           &             &         & x         & route, deadline               \\
\citeauthor{larson2013coordinated}\cite{larson2013coordinated,larson2015distributed} & & x & & x  & x           &             &         & x         & speed, position, destination  \\
\textcite{krupitzer2018towards}                         &             & x        & x      &       & x           &             &         & x         & route, speed                  \\
\textcite{burov2020platoon}                             &             & x        &        & x     & x           &             & x       &           & speed, position, platoon size \\
\textcite{zhu2022hindrance-aware}                       &             & x        &        & x     & x           &             & x       &           & speed, fuel, safety           \\
\textcite{khan2005convoy}                               &             & x        &        & x     &             & x           &         & x         & speed, fuel                   \\
\citeauthor{dao2013strategy}\cite{dao2008decentralized,dao2013strategy} & x &    &        & x     &             & x           &         & x         & destination                   \\
\textcite{hobert2012study}                              &             & x        &        & x     &             & x           &         & x         & destination                   \\
\textcite{sharma2022potent}                             &             & x        &        & x     &             & x           &         &           & route                         \\
\textcite{dokur2022edge}                                &             & x        &        & x     &             & x           &         & x         & route, destination            \\
\textcite{su2016autonomous}                             &             & x        &        & x     &             & x           &         & x         & speed, position               \\
our earlier work~\cite{heinovski2018platoon}            &             & x        & x      & x     & x           & x           &         & x         & speed, position               \\
\hline
\textbf{this work}                                      & (possible)  & x        & x      & x     & x           & x           & x       & x         & speed, position, preferred speed window \\
\hline
\end{tabular}
\end{table*}

\noindent
The previous strategies for platoon formation show that optimal groupings substantially improve the performance gain.
Many studies are targeted towards trucks using known transport assignments and deadlines.
We target individual vehicles, where vehicles' properties are not known beforehand and are quite diverse (e.g., variation in desired driving speed).
Thus, \vtp{} assignments need to be computed on-demand and en route, based on similarity (e.g., in driving speed) among individual vehicles instead of total cost (such as fuel consumption).
Since a vehicles' platoon choice can have a huge impact on its platooning benefits, assignments should be optimal, resulting in an NP-hard optimization problem.
Existing studies consider many different (sets of) optimization objectives and considered properties, often focusing on system-level aspects instead of vehicles' individual requirements.
While some studies perform a comparison of their centralized optimal solution with a centralized heuristic, a proper comparison between centralized and distributed approaches as well as optimal and greedy solutions is still missing from the literature.
We summarize the existing studies with on-demand \vtp{} assignment computation from this \lcnamecref{sec:related} in \cref{tab:rw}.

In earlier work~\cite{heinovski2018platoon}, we studied en route platoon formation for individual passenger vehicles based on their similarity in desired driving speed and position on the road, thereby optimizing \vtp{} assignments regarding individual properties.
Besides formally describing the platoon formation problem, we introduced both a centralized and a distributed greedy heuristic for solving the NP-hard assignment problem in an efficient but simple way.
Simulations already indicated that the choice of the approach, and the willingness to deviate from individual objectives, have a huge impact on the resulting assignments.
However, the absolute performance of the approaches in comparison to an optimal solution for all vehicles was still unclear.

In this paper, we go one step further and add an approach to optimally solve our similarity-based optimization problem using a \ac{MIP} solver.
For this, we refined the entire problem formulation to make it more applicable for centralized optimization.
We compare both greedy approaches to the optimal solution in detail using a wide range of performance metrics in a large-scale simulation study. 

%

\section{Computing \vtp{} Assignments}%
\label{sec:platoon_formation}

\noindent
In this \lcnamecref{sec:platoon_formation}, we define the problem of computing \vtp{} assignments, which has to be solved as part of the platoon formation process.
We describe our perspective on this problem as well as our approach for solving the problem using the concept of similarity among vehicles, which we base on their individual properties.

\subsection{Assumptions}%
\label{sec:assumptions}

\noindent
We focus on individual traffic, where vehicles have different properties (such as driving speed) and start their trips unsynchronized.%
\footnote{%
We assume drivers can freely customize their trip based on an individually preferred traveling speed during trip planning.
This can be compared with choosing a route from multiple options such as \emph{fast} or \emph{economic} that are provided by the navigational systems already today.%
}

Thus, the trips and requirements of vehicles are not known beforehand, which requires dynamic on-demand computation of assignments based on the current situation.
Vehicles will initially drive individually until they encounter an appropriate (existing) platoon to join or an individual vehicle to form a new platoon with.
We generally assume that driving in a platoon is preferred because of the expected benefits such as driving efficiency as well as increased safety and traffic throughput.

When vehicles start their trip (outside of the freeway), they have no knowledge about other already existing vehicles and vice versa.
We assume that the entire platoon formation process including the computation of \vtp{} assignments happens during the vehicles' trips on the freeway (en route).
%
%
After assignment, vehicles create a new platoon or join an existing one by performing a maneuver.
Vehicles can communicate with each other by means of 5G-based \ac{C-V2X} or \ac{DSRC} to exchange platooning information and maneuver control.
As platoon coordination has low requirements on the communication, we neglect this part in the evaluation to make the approach also technology agnostic.
After successful competition of the join maneuver, vehicles stay within the platoon until they reach their destination, at which a leave maneuver is performed.
%
Assuming a fully operational \aca{CACC}, vehicles in a platoon always mirror the behavior of the platoon leader and keep a constant gap of \SI{5}{\m}~\cite{rajamani2000demonstration,jootel2012final}.

\subsection{Problem Formulation}%
\label{sec:problem_formulation}

\noindent
Early approaches to find candidate vehicles to construct platoons consider different constraints and optimization goals, such as grouping by destination or route, or (pairwise) by fuel efficiency.
To avoid assigning vehicles to platoons with heterogeneous properties, we aim to consider the individual vehicles' properties during the assignment computation.
Thus, the platooning benefits for the individual vehicles can be optimized without giving up too much of the own requirements instead of system-level aspects such as traffic flow.
Assigning vehicles to platoons based on their properties is like clustering vehicles according to some similarity metric corresponding to the constraints and goals introduced by a formation strategy.

Inspired by our earlier work~\cite{heinovski2018platoon}, we are using the desired driving speed as a primary similarity metric.
Additionally, we also take the position of the vehicles on the freeway into account.
Since joining a platoon which is far away can add a lot of overhead in fuel-consumption, it not useful to consider such cases as possible options.

In order to come up with a formation strategy, we formalize the problem as follows.
By considering individual vehicles as platoons of size \num{1}, we can model both similarly by the set
\begin{equation}\label{eq:property_set}
    \{n, D, p, l \} \text{~,}
\end{equation}
where
$n$ is a unique identifier,
$D$ is the desired driving speed,
$p$ is the current position,
and $l$ is the current position of the last vehicle in the platoon or $l := p$ if vehicle $n$ is driving individually.
The position of a vehicle is defined as the location of its front bumper on the freeway.
In case of a platoon, we consider its lead vehicle (leader) as the platoon's representative.
Thus, the leader's values are used for $n$, $D$, and $p$.

We define $C$ as the set of all individual vehicles that need to be assigned to a platoon.
Additionally, we define $T$ as the set of all already existing platoons (consisting of multiple vehicles) as well as individual vehicles, representing all potential assignment targets.
Thus, all vehicles in $C$ are also included in $T$.
We define $I$ and $J$ as index sets of $C$ and $T$, thus $c_{i} \in C$ and $t_{j} \in T$ define the $i$-th and $j$-th element of $C$ and $T$, respectively.

We define $A$ as a $|C| \times |T|$ matrix that contains a decision variable $a_{ij}$ for every possible \vtp{} assignment between elements of $C$ and $T$.
The decision variable is used to determine whether an arbitrary $c_i \in C$ is assigned to an arbitrary $t_j \in T$, using the definition of $a_{ij}$ as
\begin{equation}\label{eq:decision_variable}
{a_{ij}} =
\begin{cases}
    {1,} & {\text{if }}\ c_i \text{~is assigned to~} t_j\\
    {0,} & {\text{otherwise}}
\end{cases}%
.
\end{equation}
If a searching vehicle $c_i$ cannot be assigned to any other vehicle or platoon $t_j$ (with $c_i \neq t_j$), it will need to keep driving individually.
We model such a case by $a_{ij} = 1$ with $c_i = t_j$ and call the assignment a \emph{self-assignment} of vehicle $c_i$.
Thereby, all searching vehicles $c_i$ within the assignment matrix $A$ have an assignment.
It is important to note that $A$ in general is not symmetric and not all assignments of $c_i$ and $t_j$ are technically possible (see below).

We also define $F$ as a $|C| \times |T|$ matrix that contains the similarity (or rather the deviation) $f_{ij} = f\left(c_i,t_j\right)$ in their properties between elements of $C$ and $T$, using the definition of $f\left(c,t\right)$ as
\begin{equation}\label{eq:function_f}
    f\left(c,t\right) = \alpha \cdot d_{s}\left(c,t\right) + (1 - \alpha) \cdot d_{p}\left(c,t\right)
    \text{~,}
\end{equation}
where, $\alpha \in [0,1]$ is a weighting coefficient.
The deviation between $c_i$ and $t_j$ in desired speed $d_{s}\left(c_i,t_j\right)$ as well as in position on the freeway $d_{p}\left(c_i,t_j\right)$ are defined as
\begin{align}
    d_{s}\left(c,t\right) = \frac{\| D_{c} - D_{t} \|}{m \cdot D_{c}}, \quad m \in [0,1]%
    \label{eq:function_ds}
    \text{~,}
    \\
    d_{p}\left(c,t\right) =
    \frac{\text{min} \left( \| p_{c} - p_{t} \|, \| l_{t} - p_{c} \| \right)}{r}, \quad r \in \mathbb{N}%
    \label{eq:function_dp}
    \text{~,}
\end{align}
where $m$ is the maximum allowed deviation from the desired driving speed of vehicle $c$
and $r$ is the maximum allowed deviation from the position on the freeway of vehicle $c$ (i.e., the search range) for vehicles being considered as potential candidates.
\Cref{eq:function_ds,eq:function_dp} both calculate a deviation relative to the respective maximum value (i.e., $m$ and $r$), thereby defining a window of allowed deviation in $[0,1]$.
While \cref{eq:function_ds} simply uses the desired driving speed of $c$ and $t$, \cref{eq:function_dp} considers the location of vehicle or platoon $t$ in relation to vehicle $c$.
We assign the largest possible deviation value (i.e., $1.0$) for modeling the similarity of a vehicle to itself ($c_i = t_j$).
Thereby, we allow a self-assignment, but clearly favor assignments to other vehicles or platoons.

Typical approaches for calculating user similarity include the Jaccard coefficient, the cosine similarity, and the multi-dimensional Euclidean distance~\cite{wu2019energy-efficient}.
These, however, cannot be applied to our problem as they either work only on sets (Jaccard) or do not support defining a maximum deviation as well as importance per property (cosine and Euclidean).
Thus, our model uses a one-dimensional Euclidean distance that is normalized based on a maximum allowed deviation (a deviation window) and weighted using a coefficient per property.
\Cref{eq:function_f} can flexibly be extended with arbitrary other properties while keeping a total value in $[0,1]$.

The previous equations are subject to the following constraints:
\begin{align}
    d_{s}\left(c,t\right) \leq 1.0%
    \label{eq:constraint_window}
    \text{~,}
    \\
    d_{p}\left(c,t\right) \leq 1.0%
    \label{eq:constraint_range}
    \text{~,}
    \\
    p_{c} \leq l_{t}%
    \label{eq:constraint_position}
    \text{~,}
\end{align}
where $p_{c}$ is the position of vehicle $c$ and $l_{t}$ is the position of the last vehicle in the platoon $t$ ($l_{t} = p_{c}$ if $t$ is driving individually).
It is important to mention that \cref{eq:constraint_position} requires that a vehicle or a platoon $t$ is in front of the searching vehicle $c$ in order to be considered as a potential candidate.
Thereby, without loss of generality, we currently only allow joining at the back of a vehicle or platoon, which requires less coordination~\cite[98]{hobert2012study} and does not influence other vehicles as well as already existing platoon members as much (assuming only the joining vehicle adjusts its driving speed in order to approach the platoon).
``If an existing platoon [had] to slow down to allow a following single vehicle to join (faster), there would be significant wasted energy as multiple vehicles [had] to undergo braking and subsequent acceleration.''~\cite[7574]{liu2022optimizing}.
Considering also vehicles in the front can increase the probability of finding a candidate and will potentially influence the role the vehicles will have in the resulting platoon (leader vs.\ follower).
For conciseness, we do not consider those effects in this work.
It is important to note that, like $A$, also $F$ in general is not symmetric.

Using the previous definitions, we can describe the objective of our optimization problem as
\begin{equation}\label{eq:optimization_problem}
    \text{minimize} \sum_{i \in I}{\sum_{j \in J}{a_{ij} f_{ij}}}
    \text{~,}
\end{equation}
subject to the following constraints
\begin{align}
\forall i \in I : \sum_{j \in J}{a_{ij}} = 1%
\label{eq:constraint_exactly_one_assignment}
\text{~,}
\\
%
\forall j \in J : \sum_{i \in I}{a_{ij}} \leq 1%
\label{eq:constraint_at_most_one_assignment}
\text{~,}
\end{align}
\begin{equation}\label{eq:constraint_only_assignment_if_no_other}
\begin{split}
\forall i \in I : a_{ij} = 1
\land c_i = t_j
,
\\
\text{if~}
\exists k \in I : {a_{kl}} = 1
\land c_i = t_l
\land c_k \neq t_j
\end{split}
\text{~.}
\end{equation}

To summarize, we try to find the best fitting platoon candidate (i.e., the candidate with the biggest similarity, or, in other words, with the smallest deviation) $t_j$ for each vehicle $c_i$ to form a platoon with.
\Cref{eq:constraint_exactly_one_assignment} assures that vehicle $c_i$ is assigned to exactly one target vehicle or platoon $t_j$.
Here, a self-assignment indicates that the vehicle continues to drive individually.
\Cref{eq:constraint_at_most_one_assignment} assures that at most one vehicle is assignment to every target vehicle or platoon $t_j$.
This is related to the join maneuver that needs to be executed in order to successfully implement an assignment.
\Cref{eq:constraint_only_assignment_if_no_other} assures that vehicle $c_i$ is not assigned to any vehicle or platoon $t_j$ (i.e., a self-assignment) if another vehicle $c_k$ ($\neq t_j$) is already assigned to it ($t_l = c_i$).

After assignment ($a_{ij} = 1$), vehicle $c_i$ performs a join maneuver towards its designated target platoon or vehicle $t_j$.
Assuming a successful join maneuver, vehicle $c_i$ became part of a platoon with vehicle $t_j$ (or the platoon $t_j$ was already part of).
Once vehicles become platoon members, they stay in the platoon until they reach their destination.

\subsection{Example Scenario}%
\label{sec:example_scenario}

\begin{figure}[!t]
    \centering
    \includegraphics[width=\columnwidth]{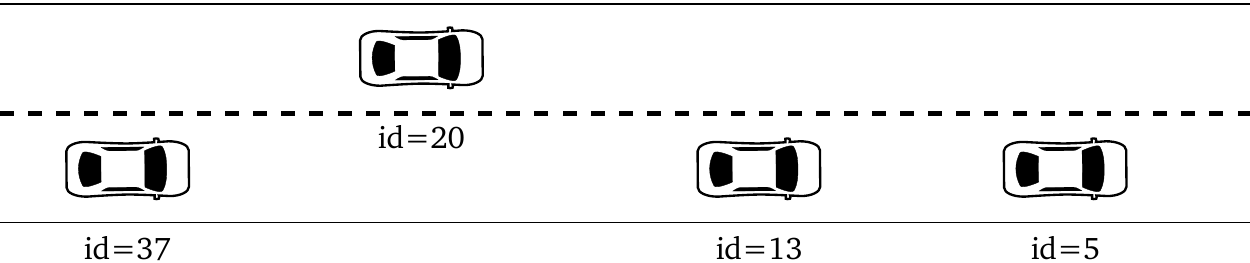}
    \caption{Example scenario: Four vehicles are driving individually on an arbitrary road with two lanes (e.g., a freeway) and try to find a platoon.}%
    \label{fig:example_scenario}
\end{figure}

\noindent
As an example for the optimization problem, consider the scenario depicted in \cref{fig:example_scenario}, where four vehicles are driving individually on an arbitrary road with two lanes (e.g., a freeway) and now try to find a platoon.
The vehicles in the example are defined by their set of properties,
$\{5, \SI{121}{\kmh}, \SI{430}{\m}, \SI{430}{\m} \}$,
$\{13, \SI{89}{\kmh}, \SI{270}{\m}, \SI{270}{\m} \}$,
$\{20, \SI{107}{\kmh}, \SI{250}{\m}, \SI{250}{\m} \}$,
$\{37, \SI{93}{\kmh}, \SI{70}{\m}, \SI{70}{\m} \}$,
and the algorithm in this case uses the following set of parameters:
\begin{equation*}
    \alpha = 0.6,
    m = 0.4,
    r = \SI{400}{\m}
    \text{~.}
\end{equation*}

By using these properties and parameters, the list of possible platoon candidates and their corresponding deviation $f\left(c,t\right)$ can be calculated as
\begin{equation}\label{eq:example_possible_assignments}
\begin{split}
    %
    & f\left(13,5\right) = 0.6 \cdot 0.599 + 0.4 \cdot 0.4 = 0.519 \\
    \\[-1em]
    & f\left(20,5\right) = 0.6 \cdot 0.218 + 0.4 \cdot 0.45 = 0.31 \\
    & f\left(20,13\right) = 0.6 \cdot 0.28 + 0.4 \cdot 0.05 = 0.188 \\
    \\[-1em]
    & f\left(37,5\right) = 0.6 \cdot 0.501 + 0.4 \cdot 0.9 = 0.66 \\
    & f\left(37,13\right) = 0.6 \cdot 0.071 + 0.4 \cdot 0.5 = 0.242 \\
    & f\left(37,20\right) = 0.6 \cdot 0.25 + 0.4 \cdot 0.45 = 0.33 \text{~.}
\end{split}
\end{equation}
It is important to note that we skipped all entries for self-assignments in this list.
By also considering self-assignments, all vehicles have at least one platoon candidate to \emph{join} with a deviation value of $\leq 1.0$.
In particular \num{5} can only be assigned to itself.

From the list of possible candidates and their corresponding deviations, the (optimal) solution minimizing the overall deviation is
\begin{equation*}
\begin{split}
    & f\left(5,5\right) = 1.0 \\
    & f\left(37,13\right) = 0.242 \\
    & f\left(20,5\right) = 0.31 \\
\end{split}
\end{equation*}
with a total deviation of $1.0 + 0.242 + 0.31 = 1.552$.
Here, selecting a candidate pair blocks both involved vehicles, making them unavailable for further selection.
Since a vehicle can only be in one maneuver at a time, at most two assignments can be executed in parallel.
After these maneuvers are finished, the vehicles in the scenario are grouped into two platoons: $\{ 13, 37 \}$ and $\{ 5, 20 \}$.

%

\subsection{Solution approach: \emph{``\optimal{}}''}%
\label{sec:approach_optimal}

\noindent
We aim to minimize the overall deviation from the desired metrics while assigning as many vehicles to platoons as possible.
Thus, in this approach, the optimization problem is solved periodically for every vehicle in the scenario at the same time.
A central server has global knowledge about all vehicles and their corresponding properties, which is collected by means of an infrastructure-based network such as 5G-based \ac{C-V2X}.
Using the vehicles' information, a mathematical solver can compute an optimal solution for all vehicles.
The complexity of centralized optimization in general has been shown to be NP-hard~\cite{larsson2015vehicle}, which makes a real-world deployment difficult, but it can provide an upper bound for possible solutions.

\begin{algorithm}
    \caption{Creation of decision variables for the solver in the \optimal{} approach}%
    \label{alg:optimal_decision_variables}
    \begin{algorithmic}
        \REQUIRE{list of all vehicles in the scenario}
        \FORALL{vehicles $c_i$ in the scenario;}
            \IF{$c_i$ in platoon \OR{} in maneuver}
                \STATE{next;}
                \COMMENT{$c_i$ is (currently) not available}
            \ENDIF{}
            \STATE{add constraint: only one assignment of $c_i$;}
            \FORALL{vehicles / platoons $t_j$ in the scenario}
                \IF{($t_j$ in platoon \AND{} \NOT{} platoon leader) \OR{} $t_j$ in maneuver}
                    \STATE{next;}
                    \COMMENT{$t_j$ is (currently) not available}
                \ENDIF{}
                \IF{$p_{c_i} > l_{t_j}$}
                    \STATE{next;}
                    \COMMENT{do not consider vehicles / platoons behind}
                \ENDIF{}
                \IF{$d_{s} \left(c_i, t_j \right) > 1.0$ \OR{} $d_{p} \left(c_i, t_j \right) > 1.0$}
                    \STATE{next;}
                    \COMMENT{too large deviation in speed or position}
                \ENDIF{}
                \IF{$c_i = t_j$}
                    \STATE{$f_{ij} = 1.0$;}
                    \COMMENT{use large deviation for self-assignment}
                \ELSE{}
                    \STATE{calculate $f_{ij}$ based on $d_{s}$ and $d_{p}$;}
                \ENDIF{}
                \STATE{add decision variable and deviation $\{ a_{ij}, f_{ij} \}$;}
            \ENDFOR{}
        \ENDFOR{}
        \ENSURE{all decision variables and deviations: list$\left( \{ a_{ij}, f_{ij} \} \right)$}
    \end{algorithmic}
\end{algorithm}

\begin{algorithm}
    \caption{Addition of constraints for the solver in the \optimal{} approach}%
    \label{alg:optimal_constraints}
    \begin{algorithmic}
        \REQUIRE{all decision variables and deviations: list$\left( \{ a_{ij}, f_{ij} \} \right)$}
        \FORALL{(unique) vehicles / platoons $t_j$ in all entries in list}
            \FORALL{entries in list with target $t_j$ and vehicle $c_i \neq t_j$}
                \STATE{add constraint: at most one assignment to $t_j$;}
                \STATE{add constraint: at most one assignment of $c_i$;}
            \ENDFOR{}
            \FORALL{entries in list with vehicle $t_j$ and target $c_i \neq t_j$}
                \STATE{add constraint: at most one assignment of $t_j$;}
            \ENDFOR{}
        \ENDFOR{}
    \end{algorithmic}
\end{algorithm}

For all possible \vtp{} assignments between $c_i \in C$ and $t_j \in T$, the decision variable $a_{ij}$ (see \cref{eq:decision_variable}) needs to be created.
This also includes vehicles that cannot be assigned (at the time being).
We show the process of creating the decision variables in \cref{alg:optimal_decision_variables}.
Note that not all combinations of $c_i$ and $t_j$ are technically possible (see \cref{eq:constraint_window,eq:constraint_range,eq:constraint_position}).
In such cases, no decision variable $a_{ij}$ is created.
Besides the list of decision variables and corresponding deviations, the solver needs some further constraints in order to compute a useful solution (see \cref{eq:constraint_exactly_one_assignment,eq:constraint_at_most_one_assignment,eq:constraint_only_assignment_if_no_other}).
We depict the process of adding the constraints to the model for the solver in \cref{alg:optimal_constraints}.
After the solver has computed a solution, it has to be applied to the vehicles.
The solution is stored within the decision variables of the model, which are used to trigger corresponding join maneuvers among involved vehicles.
Due to the centralized coordination, all assignments are synchronized, and maneuver conflicts are thereby avoided.
In case of a self-assignment, no maneuver is triggered, and the vehicle stays individual.

\subsection{Solution approach: \emph{``\centralized{}}''}%
\label{sec:approach_centralized}

\noindent
In order to make a real-world deployment of a centralized solution for the optimization problem more feasible, we aim to reduce its computational complexity.
Thus, we propose the \centralized{} approach, which consists of greedy heuristics following the idea of the \optimal{} approach:
it computes \vtp{} assignments for every vehicle in the scenario at the same time using full knowledge about all vehicles.

\begin{algorithm}
    \caption{Collection of possible candidates in the \centralized{} approach}%
    \label{alg:heuristic_candidates_centralized}
    \begin{algorithmic}
        \REQUIRE{all vehicles in the scenario}
        \FORALL{vehicles $c_i$ in the scenario;}
            \IF{$c_i$ in platoon \OR{} $c_i$ in maneuver}
                \STATE{next;}
                \COMMENT{$c_i$ is (currently) not available}
            \ENDIF{}
            \FORALL{vehicles / platoons $t_j$ in the scenario with $t_j \neq c_i$}
                \IF{($t_j$ in platoon \AND{} $t_j$ \NOT{} platoon leader) \OR{} $t_j$ in maneuver}
                    \STATE{next;}
                    \COMMENT{$t_j$ is (currently) not available}
                \ENDIF{}
                \IF{$p_{c_i} > l_{t_j}$}
                    \STATE{next;}
                    \COMMENT{do not consider vehicles / platoons behind}
                \ENDIF{}
                \IF{$d_{s} \left(c_i, t_j \right) > 1.0$ \OR{} $d_{p} \left(c_i, t_j \right) > 1.0$}
                    \STATE{next;}
                    \COMMENT{too large deviation in speed or position}
                \ENDIF{}
                \STATE{add $\{ c_i, t_j, f \left(c_i, t_j \right) \}$ to list of possible assignments;}
            \ENDFOR{}
        \ENDFOR{}
        \ENSURE{all possible assignments: list$\left( \{ c_i, t_j, f \left(c_i, t_j \right) \} \right)$}
    \end{algorithmic}
\end{algorithm}

In this approach, we first calculate the deviation $f \left(c_i, t_j \right)$ for all vehicles in the neighborhood which do not violate the constraints given in \cref{eq:constraint_window,eq:constraint_range,eq:constraint_position} and add an entry for them to a list of possible assignments, using \cref{alg:heuristic_candidates_centralized}.
An entry $\{ c_i, t_j, f \left(c_i, t_j \right) \}$ in this list contains vehicle $c_i \in C$ and vehicle or platoon $t_j \in T$ as well as the total deviation of vehicle or platoon $t_j$ from vehicle $c_i$ (see \cref{eq:function_f}).
Note that the deviation of $t_j$ from $c_i$ is not symmetric as the opposite direction produces a different value.

\begin{algorithm}
    \caption{Selection of best candidate for assignment in the \centralized{} approach}%
    \label{alg:heuristic_selection_centralized}
    \begin{algorithmic}
        \REQUIRE{all possible assignments: list$\left( \{ c_i, t_j, f \left(c_i, t_j \right) \} \right)$}
        \FORALL{(unique) vehicles $c_i$ in the list}
            \STATE{$s \leftarrow \{t_j | t_j \in \text{list} \left( \{ c_i, t_j, f_{i} \left( x \right) \} \right) \}$;}
            \COMMENT{all possible $t_j$}
            \IF{$\| s \| > 0$}
              \STATE{$b \leftarrow \min{f \left(c_i, t_j \right), t_j \in s} \}$;}
              \COMMENT{select best candidate}
              \STATE{let $c_i$ join $b$;}
              \STATE{remove list entries with $c_i$ or $b$;}
              \COMMENT{apply constraints}
              \ENDIF{}
        \ENDFOR{}
        \ENSURE{vehicles performing a join maneuver}
    \end{algorithmic}
\end{algorithm}

Once all possible assignments and the corresponding deviations are computed, we use \cref{alg:heuristic_selection_centralized} to select the best target $t_j$ for every searching vehicle $c_i$ from this list and trigger a corresponding join maneuver.
In particular, we select the $t_j$ with the smallest deviation $f \left(c_i, t_j \right)$ and remove all entries which contain vehicles $c_i$ or $t_j$ because they became unavailable for further selection.
This heuristic is greedy as it makes decisions solely from the perspective of a searching vehicle $c_i$ without considering consequences for other searching vehicles, which could join the same vehicle or platoon $t_j$.
Thus, it does not consider all possible assignments for all vehicles during the decision making.
Due to its nested for-loop, the computational complexity of this approach is $O \left( n^2 \right)$.

Looking again at the example scenario from \cref{sec:example_scenario}, we observe that the \centralized{} heuristic also produces two platoons, but it does not necessarily compute the aforementioned optimal solution.
Instead, without loss of generality, it will select
\begin{equation*}
\begin{split}
    & f\left(5,5\right) = 1.0 \\
    & f \left(13,5\right) = 0.519 \\
    & f \left(37,20\right) = 0.33 \\
\end{split}
\end{equation*}
with a total deviation of $1.0 + 0.519 + 0.33 = 1.849$ by following the given order of possible assignments from \cref{eq:example_possible_assignments} (i.e., increasing by id).
Here, selecting a candidate pair blocks both involved vehicles, making them unavailable for further selection.
Since a vehicle can only be in one maneuver at a time, at most two assignments can be executed in parallel.
Due to the centralized coordination, all assignments are synchronized, and maneuver conflicts are thereby avoided.
After these maneuvers are finished, the vehicles in the scenario are grouped into two platoons: $\{ 5, 13 \}$ and $\{ 20, 37 \}$.

%

\subsection{Solution approach: \emph{``\distributed{}}''}%
\label{sec:approach_distributed}

\noindent
Solving the optimization problem at a centralized entity for all vehicles using global knowledge about all vehicles has advantages.
In contrast, however, it has limited scalability, introduces a single point of failure, and does not necessarily present the best design considering the management of dynamic and fast-paced entities like vehicles~\cite{ergenc2022distributed}.
Thus, in order to remove the dependency on a central entity, we propose the \distributed{} approach, which consists of greedy heuristics as well but works fully distributed:
every vehicle $c_i \in C$ tries to compute a \vtp{} assignment to a vehicle or platoon in its neighborhood $\Omega_{i} \subset T$ individually.
In contrast to both centralized approaches, vehicles in the \distributed{} case only have local knowledge of the scenario:
they know other vehicles within a certain (communication) range, assuming some general neighbor management, e.g., from regular \ac{DSRC} beacons~\cite{heinovski2018platoon}, and can directly access other vehicles' information (i.e., speed, position, platoon state, and maneuver state).
In all approaches, we assume that the information is always up to date, thus, vehicles which are not applicable anymore cannot be used as platooning opportunity.



\begin{algorithm}
    \caption{Collection of possible candidates in the \distributed{} approach}%
    \label{alg:heuristic_candidates_distributed}
    \begin{algorithmic}
        \REQUIRE{list of (available) vehicles / platoons $t_j$ in communication range of vehicle $c_i$ (i.e., $\Omega_{i})$}
        \FORALL{vehicles / platoons $t_j$ in the neighborhood $\Omega_{i}$}
            \IF{$p_{c_i} > l_{t_j}$}
                \STATE{next;}
                \COMMENT{do not consider vehicles / platoons behind}
            \ENDIF{}
            \IF{$d_{s} \left(c_i, t_j \right) > 1.0$ \OR{} $d_{p} \left(c_i, t_j \right) > 1.0$}
                \STATE{next;}
                \COMMENT{too large deviation in speed or position}
            \ENDIF{}
            \STATE{add $\{ c_i, t_j, f \left(c_i, t_j \right) \}$ to list of possible assignments;}
        \ENDFOR{}
        \ENSURE{list of possible assignments: list$\left( \{ c_i, t_j, f \left(c_i, t_j \right) \} \right)$}
    \end{algorithmic}
\end{algorithm}

Using the information about nearby vehicles, the heuristic given in \cref{alg:heuristic_candidates_distributed} is executed to prepare the list of possible assignments.
Note that the deviation of $t_j$ from $c_i$ is not symmetric as the opposite direction produces a different value.
A heuristic like \cref{alg:heuristic_selection_centralized} is used to select a candidate vehicle or platoon $t_j$ with the smallest deviation to join.
Conceptually, the same assignments as in the \centralized{} approach are selected.
However, the selection of possible candidates is limited to the restricted nature of the local knowledge and, therefore, depends on the time the heuristic is evaluated.
Also, the vehicles are not synchronized and thus perform the assignment process asynchronously, leading to assignments and maneuvers not being triggered at the same time.
The computational complexity of this approach is $O \left( n \right)$.

Looking again at the example scenario from \cref{sec:example_scenario}, we observe that the \distributed{} heuristic will trigger join attempts for the following subset of \cref{eq:example_possible_assignments}:
\begin{equation*}
\begin{split}
    & f \left(13,5\right) = 0.519 \\
    & f \left(20,13\right) = 0.188 \\
    & f \left(37,20\right) = 0.33 
    \text{~.}
\end{split}
\end{equation*}
The outcome of these join attempts depends on the order of execution defined by the trigger time at the involved vehicles (i.e., \numlist{13;20;37}).
Since a vehicle can only be in one maneuver at a time, at most two assignments can be executed in parallel.
If both \num{13} and \num{37} trigger their attempts roughly at the same time but before \num{20}, two platoons will be formed: $\{ 5, 13 \}$ and $\{ 20, 37 \}$.
Otherwise, only the first triggered attempt (e.g., from \num{20}) will lead to a successful platoon and the others will be aborted due to unavailability of the involved vehicle(s) (e.g., \num{20}).

%

\section{Evaluation}%
\label{sec:eval}

\noindent
Within this work, we evaluate the proposed platoon formation approaches in an extensive simulation study.
We do so by comparing them to each other and to two baseline approaches without platooning, using typical metrics for traffic modeling as well as platooning.
In the evaluation, we first report details of our three formation algorithms to show the impact of knowledge (local vs.\ global).
We then discuss results from a macroscopic, system-level perspective of the scenario to show the impact on the general traffic system.
Finally, we report results for typical platooning metrics from a microscopic, vehicle-level perspective of the individual vehicles.

%

\subsection{Methodology}%
\label{sec:methodology}

\begin{figure}[!t]
    \centering
    \includegraphics[width=\columnwidth]{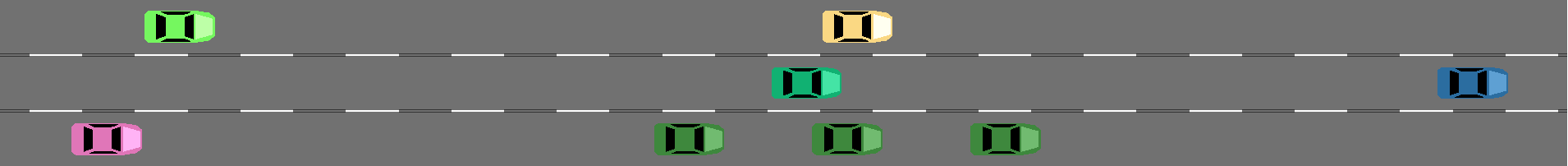}
    \caption{%
        Screenshot from the \sumo{}-based live GUI of \simulator{}, showing a small region of our simulation scenario:
        \num{3} lane freeway, \num{5} individually driving vehicles, and \num{1} formed platoon consisting of \num{3} members.
        All members of one platoon are shown in the same color, which makes it easy to recognize them.
    }%
    \label{fig:screenshot}
\end{figure}

\noindent
To observe effects of platooning and platoon formation algorithms as well as their impact, large-scale simulation studies with large scenarios and many vehicles are required.
Many studies use \plexe{}~\cite{segata2022multi-technology} for this purpose.
While being perfectly suited for fine-grained simulation of platoon controllers including the necessary wireless communication link, the approach suffers from limited scalability.
Therefore, we use \simulator{}\footnote{\url{https://www.plafosim.de}} in our study as we are more interested in large-scale effects of platoon management rather than microscopic control of the involved vehicles~\cite{heinovski2021scalable}. 
Besides validating its underlying mobility models in a comparison to \sumo{} and \plexe{}, we ran the experiments from our earlier work~\cite{heinovski2018platoon} to validate the algorithm behavior using \simulator{}, producing qualitatively equal results.
We were able to increase the scenario in size and simulation time as well as in the number of parameter configurations.
We use \simulator{} version 0.15.4 with some additional changes, which we will also integrate into its upstream version.\footnote{\url{https://github.com/heinovski/plafosim}}
A screenshot from the \sumo{}-based live GUI of \simulator{} is shown in \cref{fig:screenshot}.

\begin{table}
    \footnotesize
    \centering
    \caption{Simulation parameters for road and traffic}%
    \label{tab:params_traffic}
    \begin{tabular}{lr}
        \toprule
        Parameter                                   & Value \\
        \midrule
        Freeway length                              & \SI{100}{\km} \\
        Number of lanes                             & 3 \\
        Ramp interval                               & \SI{10}{\km} \\
        Depart positions                            & random on-ramp \\
        Arrival positions                           & random off-ramp at trip end \\
        \midrule
        Fixed trip length                           & \SI{50}{\km} \\
        Desired speed                               & $\mathcal{N} \left(\SI{120}{\kmh}, 10\% \right)$ \\
        Min.\ desired speed                         & \SI{80}{\kmh} \\
        Max.\ desired speed                         & \SI{160}{\kmh} \\
        Target density                              & \SIlist{5;10;15;20;25}{\density} \\
        Target no.\ of vehicles                      & \SIlist{1500;3000;4500;6000;7500}{\vehicle} \\
        Departure rate                              & \SIlist{3564;7129;10693;14257;17822}{\flow} \\
        \midrule
        \acl{CF} model                              & Krauss, \ac{ACC}, and \ac{CACC} \\
        Krauss desired headway                      & \SI{1}{\s} \\
        \ac{ACC} desired headway                    & \SI{1}{\s} \\
        \ac{CACC} desired gap                       & \SI{5}{\m} \\
        Max.\ speed $v_{\text{max}}$                & \SI{200}{\kmh} \\
        Max.\ acceleration                          & \SI{2.5}{\m\per\s\squared} \\
        Max.\ deceleration                          & \SI{10}{\m\per\s\squared} \\
        Vehicle length                              & \SI{5}{\m} \\
        Min.\ gap between veh.\                     & \SI{2.5}{\m} \\
        \bottomrule                                 \\
    \end{tabular}
\end{table}

In our simulation study, we consider a \num{3}-lane freeway (see \cref{fig:screenshot}) of \SI{100}{\km} length with periodic on-/off-ramps every \SI{10}{\km}, which allow vehicles to enter and leave the freeway.
Vehicles perform trips of \SI{50}{\km} between a pair of randomly selected on-/off ramps.
We assume a road network without any disturbances to the road infrastructure (e.g., by road construction) or by traffic accidents.

The desired driving speed is sampled from a normal distribution with a mean of \SI{120}{\kmh} (\SI{33}{\mps}), a variation of \num{10}\%, and limited to values from $\left[\SI{80}{\kmh}, \SI{160}{\kmh}\right]$.%
\footnote{%
The parameters roughly correspond to a typical German Autobahn with a recommended driving speed of \SI{130}{\kmh} and allow us to account for the desired driving speed (and limit) of trucks (\SI{80}{\kmh}) as well as faster driving vehicles (\SI{160}{\kmh}).
We observed only few vehicles (\num{0.4}\%) that were assigned a desired driving speed equal to the artificial limits of our distribution.%
}
Vehicles start their trips driving individually, using either the popular Krauss model~\cite{krauss1998microscopic} or a standard \ac{ACC} (cf.\ \cite[Eq.\ 6.18]{rajamani2012vehicle}).
They keep a constant time-based safety gap order to avoid collisions.
Platoon members use a standard \ac{CACC} with constant spacing (cf.\ \cite[Eq.\ 6]{rajamani2000demonstration}) after platoon formation is completed.

In our study, we compare our three proposed platoon formation algorithms with two baseline approaches without platooning:
\begin{itemize}
\item \human{} -- human driving (following the Krauss model)
\item \acc{} -- vehicles controlled by a standard \ac{ACC}
\item \distributed{} -- vehicle platooning (\ac{CACC}) using our distributed (greedy) approach (see \cref{sec:approach_distributed}) for platoon formation
\item \centralized{} -- vehicle platooning (\ac{CACC}) using our centralized (greedy) approach (see \cref{sec:approach_centralized}) for platoon formation
\item \optimal{} -- vehicle platooning (\ac{CACC}) using our optimal approach (see \cref{sec:approach_optimal}) for platoon formation
\end{itemize}
Thus, we can determine the improvement of vehicle platooning (\ac{CACC}) in comparison to already existing but non-platooning driver assistance systems.



\noindent
We model vehicle demand in a macroscopic way via a flow with constant insertion (i.e., new vehicles are continuously being inserted based on a departure rate) to roughly keep a given target density of vehicles on the road in order to achieve a stable traffic situation.
We do so because we are only interested in a relative comparison between non-platooning and platooning approaches and not in the maximum possible traffic flow for every approach.
The corresponding departure rate for a given target density is calculated as
\begin{align}
    t_\text{expected} = \frac{d_\text{trip}}{v_\text{desired}} \label{eq:typical_trip_duration}\\
    P_\text{arrival} = \frac{1}{t_\text{expected}} \label{eq:vehicles_arriving_per_second}\\
    R_\text{departure} = \left[P_\text{arrival} \cdot D_\text{desired} \cdot L \cdot d_\text{road} \cdot 3600\right] \label{eq:depart_rate}
\end{align}
where
$t_\text{expected}$ is the expected travel time of the vehicles, given their fixed trip length and their desired driving speed (in \si{\mps}),
$P_\text{arrival}$ is the probability of vehicle arrivals per second,
$R_\text{departure}$ is the corresponding departure rate, given the target density $D_\text{desired}$, number of lanes $L$, and the length of the road $d_\text{road}$.
A summary of all simulation parameters for the road network and traffic can be found in \cref{tab:params_traffic}.
We chose \SI{25}{\density} as highest target vehicle density as initial simulations showed that at higher densities the scenario is too crowded to insert further vehicles in all simulated approaches.


\begin{table}
    \footnotesize
    \centering
    \caption{Simulation parameters for formation logic}%
    \label{tab:params_formation}
    \begin{tabular}{lr}
        \toprule
        Parameter                                   & Value \\
        \midrule
        Penetration rate                            & 100\% \\
        Execution interval                          & \SI{60}{\s} \\
        Communication range for \distributed{}   & \SI{500}{\m} \\
        \midrule
        Max.\ deviation from desired speed $m$      & \numrange{0.1}{0.3}, step \num{0.1} \\
        Max.\ deviation in position $r$             & \SI{1000}{\m} \\
        Weight of speed deviation $\alpha$          & \num{0.5} \\
        Time limit for \ac{MIP} solver              & \SI{600}{\s} \\
        \bottomrule
    \end{tabular}
\end{table}

Platoon formation is performed in regular intervals every \SI{60}{\s}.
Forming platoons always consist of the following steps:
(1) data collection of available vehicles and platoons,
(2) computation of vehicle-to-platoon assignments depending on the selected approach,
(3) execution of join maneuvers to implement the computed assignment(s).
As the focus of this work is on the assignment process, platooning mobility, wireless communication, and platooning maneuvers are implemented in an abstracted way~\cite{heinovski2021scalable}.
A summary of all simulation parameters for the platoon formation can be found in \cref{tab:params_formation}.
In order to consider deviation in desired speed and position equally, we set $\alpha = 0.5$ within this study.
Further, we set the maximal allowed distance between vehicles $r = \SI{1000}{\m}$ to investigate the benefit of the centralized approaches, which are not limited by the communication rage (\SI{500}{\m}).
At the same time, it avoids assignments to far away platoon candidates that cannot be reached within a reasonable time.
In general, however, vehicles can be assigned to (and join) any other vehicle or platoon which fulfills the constraints given in \cref{sec:problem_formulation}.


The \optimal{} approach is implemented by using a \ac{MIP} solver from Google's OR-Tools library.\footnote{OR-Tools version 9.0, \url{https://developers.google.com/optimization/}}
The constraints for the assignment defined in \cref{sec:approach_optimal} are modeled by row constraints and a integer decision variable with a coefficient corresponding to the respective deviation is added for every possible \vtp{} assignment.
Since self-assignments are inherently allowed but not desired, a deviation value of $1.0$ (the maximum) is used for these cases.
To achieve a finite but useful execution time, we limit the solver's execution time to \SI{600}{\s}.
While the solver is running, the simulation time does not advance, and its state does not change.
Thus, the solver's execution time does not impact the vehicles' travel time and corresponding statistics.

During join maneuvers, vehicles are moved to their designated platoon position, considering the approximate time for approaching under perfect conditions.
For real-world deployments, solutions for coping with other vehicles interfering the maneuvers have been proposed~\cite{segata2014supporting,strunz2021coop,paranjothi2020pmcd}.
We emulate a safe but lengthy join procedure without actual simulation of the required steps, thereby reducing simulation complexity and time, while also avoiding conflicts among vehicles.
It is important to mention that we currently only allow joining at the rear of a vehicle or platoon and thus only vehicles in front are considered as potential platooning candidates (cf.\ \cref{sec:problem_formulation}).
After successful competition of the join maneuver, vehicles stay within the platoon until they reach their destination ramp, at which a leave maneuver is performed.
If the platoon leader leaves the platoon, the next remaining vehicle within the formation becomes the new leader, keeping all properties of the platoon.
If all other platoon members leave the platoon, the vehicle continues to drive individually and starts searching for new platooning opportunities.



We simulate \SI{7200}{\s} (\SI{2}{\hour}) of traffic in multiple repetitions for every configuration and approach.
We first pre-fill the road network with the desired number of vehicles before the simulation starts.
To cut off the initial transient period, we use the first \SI{1800}{\s} (\SI{0.5}{\hour}) of the simulation time as a warm-up period and ignore all results in this interval, only considering vehicles that departed after this period.
Using all approaches and parameter combinations defined in \cref{tab:params_traffic,tab:params_formation}, we simulate a total of \num{165} individual runs.
Using an AMD Ryzen 9 5900X 12-Core processor @3.7GHz, we observe the shortest simulation time (\SI{50}{\minute} wall-clock time) with the smallest configuration (speed window \num{0.1}, density \SI{5}{\density}) for the \ac{ACC} approach.
In contrast, we observe the longest simulation time (\SI{4}{\hour} wall-clock time) with the largest configuration (speed window \num{0.3}, density \SI{25}{\density}) using the \optimal{} approach.
In the following, we omit showing confidence intervals for any reported average value, but we made sure that they are reasonably small.

\subsection{Validation}%
\label{sec:eval_validation}

\begin{figure}[!t]
    \centering
    \includegraphics[width=\columnwidth]{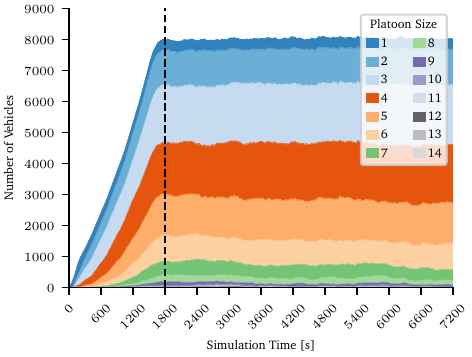}%
    \caption{%
        Total number of vehicles in platoons over the entire simulation time for the \distributed{} approach using the largest configuration (speed window \num{0.3}, density \num{25}\densityCaption{}).
        The platoon size is indicated by color and the warm-up period by the dashed line.
    }%
    \label{fig:platoon_size_life}
\end{figure}

\noindent
We performed an extensive validation of the implemented solutions.
As an example, we show the total number of vehicles in platoons for an exemplary simulation run over its entire simulation time in \cref{fig:platoon_size_life}.
Starting with \num{0} vehicles at \SI{0}{\s}, the total number of vehicles grows over time until it reaches a almost constant value of $\approx 8000$.
Slowly, platoons are formed and we can see that platoons sized of up to \num{14} vehicles are created.
After the warm-up period of \SI{1800}{\s} (dashed line), the total number of vehicles in the simulation as well the distribution of platoon sizes stays almost constant.
Other approaches and configurations show similar simulation behavior, but the total number of vehicles in the simulation as well as the distribution of platoon sizes are slightly different as we will report in the following.

%

%

%

\subsection{Available Candidates}%
\label{sec:results_candidates}

\noindent
The first step in the process of forming platoons is to find available platooning opportunities (candidates).
These can be either individual vehicles or already existing platoons.

%
\begin{figure}[!t]
    \centering
    \includegraphics[width=\columnwidth]{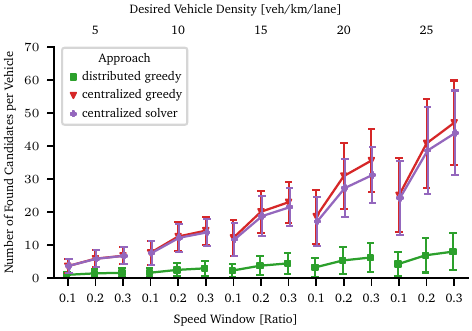}
    \caption{%
        Average number of potential platooning candidates known to the platoon formation algorithm in every iteration (with standard deviation) per vehicle and for all platooning approaches.
        The x-axis shows various values of the speed window (parameter $m$) for our platoon formation algorithms.
        The facets show various values for the desired vehicle density in the scenario.
    }%
    \label{fig:candidates_found}
\end{figure}

We show the results of the found candidates metric (i.e., the average number of potential platooning opportunities known to the platoon formation algorithm) in \cref{fig:candidates_found}.
The found candidates metric counts the number of possible assignments for platoon formation as identified
for a single vehicle.

A fundamental difference between the approaches is the available knowledge about other vehicles.
While vehicles using the \distributed{} approach only have local knowledge of other vehicles in their vicinity limited by a certain communication range, both centralized approaches have global knowledge of all vehicles within the entire scenario.
Obviously, the more vehicles are known to the respective approach, the more vehicles can be used within the formation algorithm, increasing the probability to find a possible candidate.
Of course, non-platooning approaches such as \human{} and \acc{} do not have any knowledge of other vehicles since no \ac{V2X} communication is used.
Thus, these approaches are not shown here.

As expected, both centralized approaches similarly find a lot more potential candidates to use for the platoon formation than the \distributed{} approach due to their overall knowledge of the scenario.
In fact, the \centralized{} approach finds a few more candidates than the \optimal{} approach, starting at medium densities, because the \optimal{} is able to fit more vehicles to platoons, thus, less are available for the next assignment rounds.

The number of found candidates for all approaches is also impacted by the speed window (i.e., the allowed speed deviation) as well as the desired vehicle density in the scenario.
Larger speed deviation thresholds lead to a less strict definition of similarity (in desired driving speed), which leads to more potential platooning candidates being found.
Similarly, higher desired vehicle densities lead to more vehicles in the overall scenario and thus more platooning opportunities being available in general.
This effect is more prominent for both centralized approaches as their global knowledge of the scenario allows for a significant increase in candidates with a bigger density.

The variance in values (we show the standard deviation) for all approaches is related to the desired driving speed (distribution).
Vehicles with a lower desired driving speed also have a smaller absolute (in \si{\mps}) speed window and thus less candidates.
Further, vehicles with a desired driving speed from the outer area of the normal distribution (e.g., $\leq \SI{90}{\kmh}$ and $\geq \SI{150}{\kmh}$) that we use to assign the speed, will have less candidates due to the limiting of the possible values, which are also less frequent.
In contrast, vehicles with a desired driving speed of a frequent value (e.g., \SI{120}{\kmh}) have a lot of candidates.
As expected, a larger speed window and a higher desired vehicle density increase the variance as well.
This effect is pronounced by the global knowledge of both centralized approaches.

%
\begin{figure}[!t]
    \centering
    \includegraphics[width=\columnwidth]{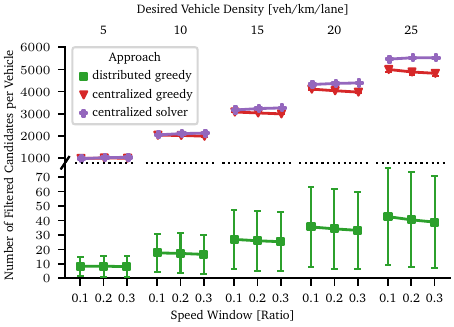}
    \caption{%
        Average number of filtered platooning candidates in every iteration (with standard deviation) per vehicle and for all platooning approaches.
        The x-axis shows various values of the speed window (parameter $m$) for our platoon formation algorithms.
        The facets show various values for the desired vehicle density in the scenario.
        The y-axis is split and does not show values between \num{80} and \num{800}.
    }%
    \label{fig:candidates_filtered}
\end{figure}

Even though the general pattern of the found candidates metric is in line with our expectations, the actual numbers are smaller than expected.
Especially both centralized approaches should find tremendously more candidates due to their global knowledge about all vehicles (cf.\ vehicle densities).
The reason for this is that some vehicles are not available (anymore) to be considered for vehicle-to-platoon assignments.
They are either already in a platoon (i.e., a follower) or currently in the process of forming a platoon (i.e., in a maneuver).
We count these vehicles within the filtered candidates metric for all vehicles and in all approaches, similarly to the found candidates metric.
Vehicles that are not available because of their position, a too large deviation, or the limited communication range with the \distributed{} approach are not counted as filtered.

The results are shown in \cref{fig:candidates_filtered}.
Due to the global knowledge and decision making for all vehicles in the scenario, both centralized approaches need to filter almost all vehicles due to their platooning status.
The \distributed{} approach needs to filter tremendously less candidates (i.e., two orders of magnitude) due to its limited local knowledge of other vehicles.
%
%
In the \optimal{} approach, more vehicles are already in a platoon and thus filtered in comparison to the \centralized{} approach due to the synchronized and balancing (leader vs.\ follower) decision making for all vehicles in the scenario.
%
For all approaches, the sum of the found and filtered candidates metrics defines a lower bound for the number of vehicles initially known to the formation algorithm.
Naturally, filtering all known vehicles that are not available (anymore) or have a too large deviation decreases the amount of effectively available candidates (the found candidates metric).

%

\subsection{Formation Process}%
\label{sec:results_execution}

\noindent
After analyzing the number of platooning opportunities, we now look at the actual formation process.

\subsubsection{Time to Platoon}%
\label{sec:platoon_time}

\begin{figure}[!t]
    \centering
    \includegraphics[width=\columnwidth]{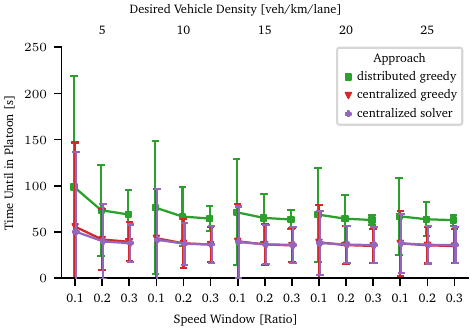}%
    \caption{%
        Average time (with standard deviation) required to become a platoon member per vehicle for all platooning approaches.
        The x-axis shows various values of the speed window (parameter $m$) for our platoon formation algorithms.
        The facets show various values for the desired vehicle density in the scenario.
    }%
    \label{fig:time_until_platoon_avg}%
\end{figure}

\noindent
\Cref{fig:time_until_platoon_avg} shows the average time required to become a platoon member per vehicle for all platooning approaches.
This time is measured from the departure of a vehicle until a join maneuver is completed successfully.
It includes waiting for (successful) execution(s) of the formation algorithm to compute a vehicle-to-platoon assignment as well as the corresponding join process itself, which in general was marginal.
Similarly, this time is measured if a vehicle became a leader of a newly formed platoon.
The actual runtime of the formation algorithm itself is not included as we did not model it for both greedy heuristics (i.e., \distributed{} and \centralized{}) and report the time for solving the optimization problem in the \optimal{} approach separately in \cref{sec:results_execution_solver}.

We immediately observe that the \distributed{} approach experiences the longest time for all configurations.
This is expected, as this approach has only limited local knowledge about available platooning opportunities.
Thus, it requires more algorithm executions until a (possible) platooning opportunity can be found (and successfully joined).
This effect becomes less prominent for bigger speed windows and larger vehicle densities.
%
Both centralized approaches require less time in general and differ only marginally.
This is expected, as both are using global knowledge about the vehicles and can thus compute successful assignments already at the first execution of the algorithm.
The standard deviations due to the variance in the number of available candidates (see \cref{fig:candidates_found}) as well as the duration between departure and execution of the assignment process.
\subsubsection{Platoon Size}%
\label{sec:platoon_size}

\noindent
Another typical metric in the context of platooning is the platoon size and its distribution.
Unfortunately, the interpretation of the platoon size is difficult and requires specific context of the underlying design goals of an algorithm.
In general, however, we aim at a low number of vehicles that are not in a platoon.


\begin{figure*}[!t]
    \centering
    \includegraphics[width=\textwidth]{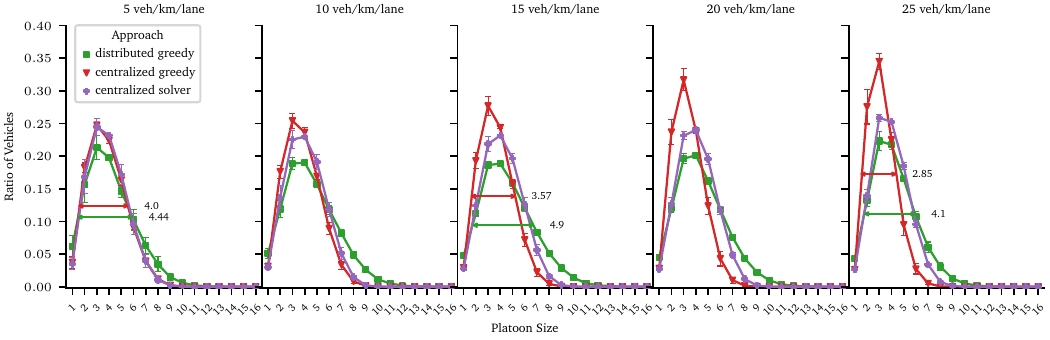}
    \caption{%
        Distribution of platoon sizes depicted by the number of vehicles in platoons with their respective size for all platooning approaches.
        The facets show various values for the desired vehicle density in the scenario.
        The horizontal arrows and corresponding numbers indicate selected full width at half maximum values of the distribution with the same color.
    }%
    \label{fig:platoon_size_distribution}
\end{figure*}

We report the distribution of the platoon size over various desired traffic densities in \cref{fig:platoon_size_distribution}.
First, we observe that all platooning approaches produce platoons (platoon size $>1$) and only a few vehicles do not find a platoon.
While, the \centralized{} approach builds slightly smaller platoons than \optimal{}, the \distributed{} approach builds slightly larger ones (max.\ sizes: \numlist{10;11;16}).
For all approaches, most of the vehicles are in platoons with a size of \numrange{2}{5}.

\subsection{System-level Metrics (Macroscopic)}%
\label{sec:results_macro}

\noindent
After reporting details of the formation process itself, we now look at traffic-related metrics from a broader, macroscopic perspective.

\subsubsection{Traffic Flow}%
\label{sec:traffic_flow}

%
\begin{figure*}[!t]
    \centering
    \subfloat[Departure Flow]{%
    \includegraphics[width=\columnwidth]{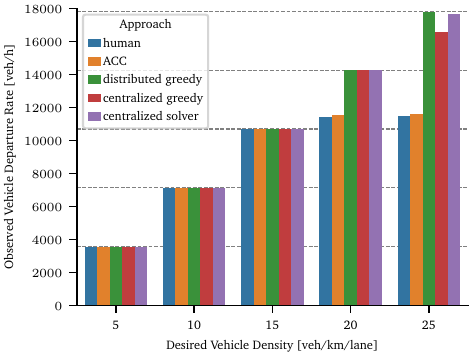}%
    \label{fig:traffic_flow}%
    }
    \hfill
    \subfloat[Vehicle Density]{%
    \includegraphics[width=\columnwidth]{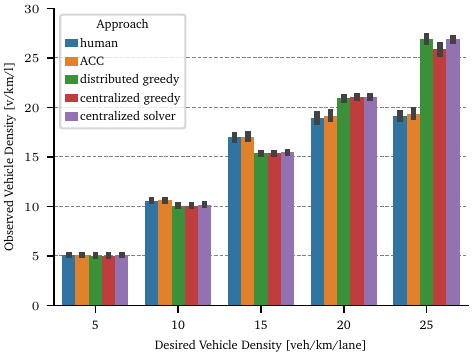}%
    \label{fig:traffic_density}%
    }
    \caption{%
        General traffic behavior observed from the simulation (with standard deviations) over all simulated approaches.
        The x-axis shows various values for the desired vehicle density in the scenario.
    }%
    \label{fig:traffic}
\end{figure*}

\noindent
Since platooning is a cooperative driving application, the performance of our algorithms and the platooning itself is influenced by the general traffic in the scenario.
All approaches will have an impact on the resulting traffic, and so does the general configuration of our simulation scenario (e.g., the departure rate).
In order to get an overview of the general traffic behavior in the scenario, \cref{fig:traffic} shows the actual departure rate (\cref{fig:traffic_flow}) as well as the resulting vehicle density (\cref{fig:traffic_density}) observed from the simulation for all of our simulated approaches and target vehicle densities.
Since both metrics are directly related to each other, the observed effects correlate as well.

The observed values for departure rate and vehicle density increase with higher desired densities.
This is consistent as we construct our departure rate from the desired vehicle density using \cref{eq:depart_rate}.
With human and \ac{ACC} driven vehicles, it is possible to meet the desired departure rates and vehicle densities at low and medium desired vehicle densities.
Starting at a desired density of \SI{20}{\density}, however, the target cannot be met anymore.
Here, the crowded traffic makes it difficult to insert new vehicles as there is only little remaining capacity (free space) on the road.
Although the platooning approaches also suffer from the same effect, they can fulfill the desired departure rates and vehicle densities for all traffic configurations and produce a smaller deviation from the desired vehicle density.


\subsubsection{Driving Speed}%
\label{sec:driving_speed}

\begin{figure}[!t]
    \centering
    \includegraphics[width=\columnwidth]{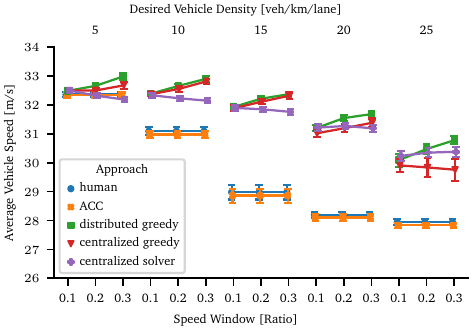}%
    \caption{%
        Average driving speed (with standard deviation) of all vehicles in the simulation for all simulated approaches.
        The x-axis shows various values of the speed window (parameter $m$) for our platoon formation algorithms.
        The facets show various values for the desired vehicle density in the scenario.
    }%
    \label{fig:driving_speed}
\end{figure}

\noindent
When looking at the average driving speed of all vehicles in the scenario (see \cref{fig:driving_speed}), we can prominently observe the effect of high traffic.
As expected, the driving speed is reduced for all approaches when the scenario becomes more crowded due to natural traffic effects.
Here, vehicles need to reduce their driving speed in order to be able to fulfill their time-based safety gap.
The effect is generally worse for non-platooning approaches due to larger safety gaps between vehicles, resulting in stop-and-go traffic~\cite{hayat2012holistic}.
For all platooning approaches, the speed is like the smallest speed window as options are limited.
But, a larger speed window allows more deviation and leads to an overall higher speed in both greedy approaches.
In comparison, the \optimal{} approach minimizes the deviation in speed and position for all vehicles simultaneously and, thus, leads to a slightly reduced driving speed.
Since larger windows allow more vehicles to become platoon leaders (instead of platoon followers), and a platoon is always driving at the speed of its leader, the deviation (in driving speed) stays small for more vehicles.
The \optimal{} approach also keeps a slightly decreasing trend over windows, being more like the original objective of vehicles.

The trend for the platooning approaches remains similar for a desired vehicle density of up to \SI{15}{\density}.
After that, the overall driving speed in the scenario decreases more noticeably for both greedy approaches than for the optimal approach;
At the largest density, despite global knowledge, the \centralized{} approach performs the worst due to synchronization and greedy selection effects.

%

%
\subsection{Vehicle-level Metrics (Microscopic)}%
\label{sec:results_micro}

\noindent
We now analyze results for platooning-related effects from a microscopic (i.e., the individual vehicles') perspective.

\subsubsection{Deviation From Desired Driving Speed}%
\label{sec:speed_deviation}

%
\begin{figure*}[!t]
    \centering
    \subfloat[%
        Results for various speed windows (parameter $m$) for a selected desired vehicle density of \num{5}\densityCaption{}.
        The results for \human{} and \acc{} are independent of the speed window.
    ]{%
    \includegraphics[width=\columnwidth]{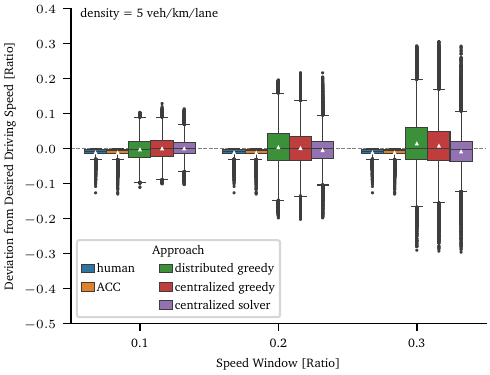}%
    \label{fig:deviation_relative_speed}%
    }
    \hfill
    \subfloat[Results for various desired vehicle densities for all simulated speed windows (parameter $m$).]{%
    \includegraphics[width=\columnwidth]{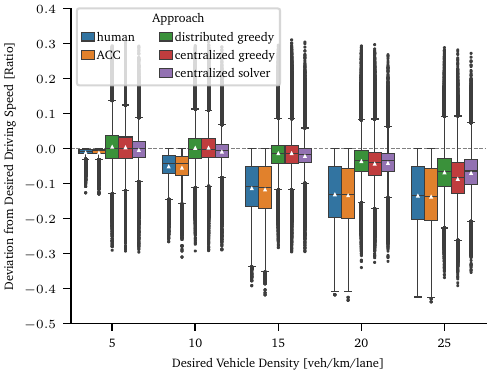}%
    \label{fig:deviation_relative_density}%
    }
    \caption{%
        Average relative deviation (ratio) from the desired driving speed per vehicle for all simulated approaches.
        The triangle within a box depicts the mean value.
    }%
    \label{fig:deviation_relative}
\end{figure*}

\noindent
Vehicles are inherently required to deviate from their desired driving speed in order to form platoons with other vehicles, which likely drive at different speeds.
The main goal of our formation algorithms is to minimize the vehicles' deviation from their own desired driving speed.
%
\Cref{fig:deviation_relative} shows the average relative deviation from the desired driving speed per vehicle for all simulated approaches.
The platooning approaches have both a negative and a positive deviation, which is due to the compromise that vehicles need to make when forming platoons (i.e., they are adjusting their driving speed to the platoon leader).
All platooning approaches perform more or less similarly and their positive deviation (driving faster than desired) mostly stays within the corresponding defined speed window (see \cref{fig:deviation_relative_speed}).

Naturally, the value of the speed window impacts the actual deviation from the desired driving speed.
As an example, we show results for a selected desired vehicle density of \SI{5}{\density} in \cref{fig:deviation_relative_speed}.
As expected, both greedy heuristics lead to a larger spread in deviation values with the \distributed{} approach having the largest.
This is due to the smaller number of platooning opportunities in this approach, which requires vehicles to join platoons with large deviations in order to fulfill the goal of platooning.
Increasing the speed window allows for even more deviation, thus also increasing the spread of values.
While the values for the \optimal{} approach are mostly symmetrically located around \num{0}, both greedy heuristics tend to deviate more positively.

Independent of platooning, the amount of vehicles in the scenario has an impact on the deviation from the desired driving speed (see \cref{fig:deviation_relative_density}).
When comparing the smallest and the largest simulated densities (\SIlist{5;25}{\density}), the negative deviation increases from nearly \num{0}\% to \num{-13}\% on average for the non-platooning approaches.
This is expected as the freeway becomes more crowded with larger densities and the vehicles need to adjust their driving speed based on the decreasing gaps to other vehicles in order to avoid crashes.
The platooning approaches cope better with a crowded scenario.
Their negative deviation reaches only \numlist{-5;-6;-9}\% on average at the largest desired vehicle density for the \distributed{}, the \optimal{}, and the \centralized{} approach, respectively.
In fact, more vehicles even lead to a slightly lower positive deviation as more platooning opportunities become available, and vehicles have to make less compromises.

%
\begin{figure}[!t]
   \centering
   \includegraphics[width=\columnwidth]{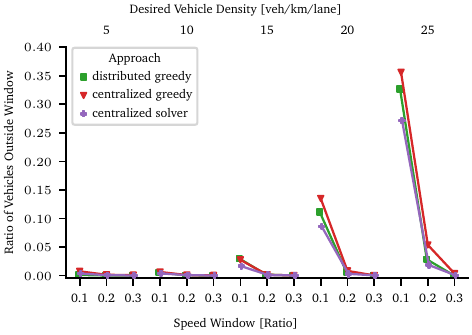}
    \caption{%
        Ratio of vehicles deviating from the allowed speed window for all platooning approaches.
        The x-axis shows various values of the speed window (parameter $m$) for our platoon formation algorithms.
        The facets show various values for the desired vehicle density in the scenario.
    }%
    \label{fig:deviation_window}
\end{figure}

In order to study the effects of the platooning approaches in more detail, we show the ratio of vehicles deviating from the allowed speed window in \cref{fig:deviation_window}.
For low densities, i.e., a small amount of available platooning opportunities, only few vehicles deviate from their desired driving speed for all approaches.
When increasing the vehicle density, the deviation increases and a difference between approaches becomes visible.
As expected, the \optimal{} approach has the lowest deviation as it tries to minimize the deviation for all vehicles in an optimal way.
With the largest speed window (\num{0.3}), all approaches perform much better because enough platooning opportunities are available to them.

\subsubsection{Travel Time}%
\label{sec:travel_time}

%
\begin{figure}[!t]
    \centering
    \includegraphics[width=\columnwidth]{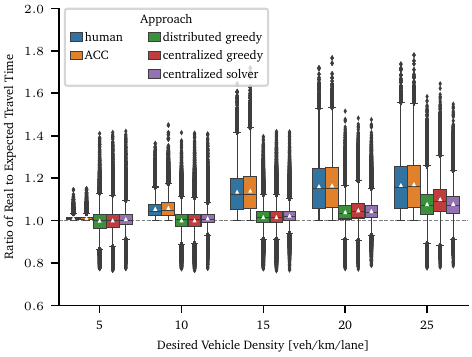}
    \caption{%
        Ratio of expected to real travel time for all simulated approaches.
        The x-axis shows various values for the desired vehicle density in the scenario.
        The triangle within a box depicts the mean value.
        A value smaller than \num{1.0} indicates that the vehicle reached its destination faster than expected, a value larger than \num{1.0} in contrast indicates that the vehicle was slower than expected.
    }%
    \label{fig:travel_time}
\end{figure}

\noindent
Naturally, the previous effects on the driving speed can be seen when looking at vehicles' trip duration as their trip lengths are equal.
When starting the trip, every vehicle estimates the time it is going to travel to its destination, assuming a constant speed at the desired value.
Upon arrival, vehicles also record their real travel time and calculate the travel time ratio (i.e., the ratio of expected to real travel time).
We use this metric to show the impact of the traffic and the use of platooning in general on vehicles' trip duration, which is a frequently used metric when assessing platooning systems.
%
We show the travel time ratio for all vehicles in \cref{fig:travel_time}.
The general tendency follows our expectations: the ratio increases with vehicle density due to slower driving (cf.\ \cref{fig:driving_speed}).
A ratio $> 1.0$ indicates that vehicles generally need longer for their trips than expected.
Both non-platooning approaches (i.e., \human{} and \acc{}) share \num{1.0} as minimum value as their trips can only get influenced negatively by delays due to traffic.
This effect is increased with higher vehicle densities as more vehicles are more impacted by the traffic.
%
Similarly, the traffic also impacts all three platooning approaches and vehicles' trips are delayed (the ratio is larger than \num{1.0} on average).
When platooning is enabled, deviations from the initially expected travel time in both directions can be observed.
Since vehicles will adjust their driving speed to the target platoon (based on the configured speed window of the formation algorithm), their speed and the resulting travel time can be faster than expected.

\subsubsection{Fuel Consumption}%
\label{sec:fuel}

\noindent
A major argument for platooning is the reduced air drag between vehicles and the resulting reduced pollutants emission and fuel consumption.
Similar results hold for electrified vehicles, we fall back to gas consumption simply because of the availability of validated and widely accepted consumption models.
%
For modeling general vehicle emissions, we use the \ac{HBEFA}\footnote{\url{https://www.hbefa.net/e/index.html}} version 3.1, following the approach implemented in \sumo{}.
In particular, we currently use the \textit{PC\_G\_EU4} emission class, which represents a gasoline driven passenger car with an engine corresponding to the European norm version 4.
For vehicles that are driving in a platoon, we calculate a reduction in emissions and fuel consumption based on their position within the platoon~\cite{bruneau2017flow}, applying the approach from our earlier work~\cite{heinovski2018platoon}.

\begin{figure}[!t]
    \centering
    \includegraphics[width=\columnwidth]{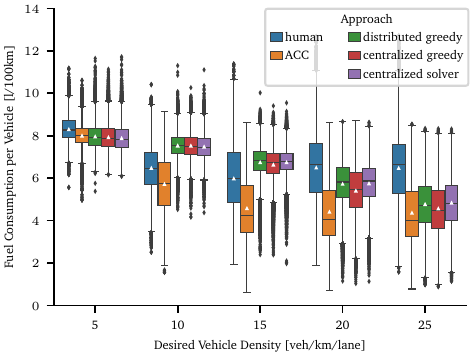}%
    \caption{%
        Fuel consumption per vehicle for all simulated approaches.
        The x-axis shows various values for the desired vehicle density in the scenario.
        The triangle within a box depicts the mean value.
    }%
    \label{fig:fuel_consumption}
\end{figure}

We show the fuel consumption in \cref{fig:fuel_consumption}.
The values plotted represent the fuel consumption of all vehicles in liters per \SI{100}{\km}, which is computed by using the total fuel consumption and the trip distance of the vehicles.
%
Following the effects shown in \cref{fig:driving_speed}, a higher target vehicle density in general reduces the fuel consumption due to the slower driving speed with more traffic in the scenario.
For human driving, starting at \SI{20}{\density}, the traffic transforms into stop-and-go, which leads to a lot of acceleration \& deceleration maneuvers increasing the average value as well as the spread of the fuel consumption.
Synchronized driving using \acc{} already helps reducing fuel consumption, in fact resulting in the lowest values, but also suffers from slower driving speed in general (cf.\ \cref{fig:driving_speed}).
All platooning approaches result in similar values, but with increasing densities, platooning seemingly performs slightly worse than \acc{}.
This is due to the higher average speed that is achieved with platooning.

\subsection{Performance of the \acs*{MIP} Solver}%
\label{sec:results_execution_solver}

%
\begin{figure}[!t]
   \centering
   \includegraphics[width=\columnwidth]{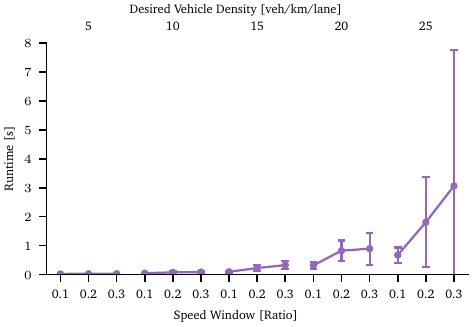}
    \caption{%
        Average runtime (with standard deviation) of all executions of the \ac{MIP} solver in the \optimal{} approach.
        The x-axis shows various values of the speed window (parameter $m$) for our platoon formation algorithms.
        The facets show various values for the desired vehicle density in the scenario.
    }%
    \label{fig:solver_runtime}
\end{figure}

\noindent
We finally look at the performance of the \ac{MIP} solver for the \optimal{} approach.
\Cref{fig:solver_runtime} shows the average runtime (with standard deviation) of all executions the solver.
This parameter measures the time that the solver uses to solve the optimization problem after all decision variables and constraints are defined (see \cref{alg:optimal_decision_variables,alg:optimal_constraints}).
As expected, the runtime increases with a larger speed window and a higher vehicle density as the number of vehicles that the solver needs to consider as source and destination increases.
It should be noted that in all configurations, the solution quality reported by the \ac{MIP} solver was well above \num{99.99}\% in all iterations.

%

\section{Discussion}%
\label{sec:discussion}


\noindent
Obviously, both centralized approaches (\optimal{} and \centralized{}) can rely on more knowledge about the scenario compared to the \distributed{} approach.
They are aware of many more vehicles and, thus, more platooning opportunities, even though many vehicles are filtered due their current platooning and maneuver status.
In the \centralized{} approach, many initially available candidates are naturally removed due to the greedy selection in the assignment process, leading to only few platooning opportunities in reality.
More vehicles in the scenario and a more relaxed speed window lead to more available candidates (i.e., platooning opportunities) in all approaches.
More platooning opportunities lead to more assignments to close-by vehicles and platoons since the similarity function in \cref{eq:function_f} also considers distance between vehicles.

The more vehicles the \ac{MIP} solver of the \optimal{} approach has to consider, the more complex the decision making becomes.
Thus, the time required for computing a solution increases while the quality of the solution decreases, given a fixed time limit for the computation.
Still, the quality of the computed solution reported by the \ac{MIP} solver was well above \num{99.9}\% in all cases, even with the largest number of vehicles and, thus, possible \vtp{} assignments that the solver has to consider.
The \optimal{} approach, however, assumes global knowledge and requires a complex \ac{MIP} solver to compute \vtp{} assignments.
In comparison, both greedy approaches have a lower computational complexity (only $O \left( n \right)$ for the \distributed{} approach), which improves scalability for real-world deployments.
Additionally, the \distributed{} approach requires only local knowledge, which eliminates the communication bottleneck towards a single (edge) server.

In all approaches, some time is needed to find a suitable platoon.
This is particularly true for the \distributed{} approach because limited local knowledge allows for less platooning opportunities.
The difference between the approaches becomes smaller with higher vehicle densities and larger allowed speed deviation.
Having only one central entity to perform the formation tasks in both centralized approaches works well for small scenarios or freeway sections, but does not scale to handle larger freeway networks (e.g., country-wide).


Obviously, the traffic density influences the driving speed and can lead to stop-and-go traffic in crowded scenarios~\cite{hayat2012holistic}.
All of our platooning approaches help improving the traffic flow by synchronizing driving.
Due to more vehicles in platoons, the \optimal{} and \distributed{} approaches perform slightly better than the \centralized{} approach.

Considering our main optimization goal, reducing the deviation from the individual desired driving speed, we observed a correlation to the traffic density:
The higher the traffic, the more the deviation from the desired driving speed for all approaches.
Platooning generally helps to reduce the deviation.
Here, \centralized{} performs slightly worse.
Adding a speed window for limiting the maximum allowed deviation, leads to reduced actual deviation due to platooning but can lead to more impact of the traffic on the driving:
\num{20}\% allowed speed deviation already helps for more than \num{90}\% of the vehicles to stay within their window.

We also looked at the fuel consumption, which is directly impacted by the traffic as well as the resulting driving speed.
A crowded scenario initially leads to less fuel due to the slower driving speed, but stop-and-go traffic increases the consumption.
Platooning helps by synchronizing the traffic and reducing the air drag of vehicles, thus saving fuel.
We see that \optimal{} and \distributed{} lead to slightly higher fuel consumption due to higher driving speed.

The deviation from the desired driving speed directly impacts the travel time.
While more traffic leads to longer trip duration, platooning can help to reduce this effect.
Again, the \centralized{} approach performs slightly worse.

From all experiments we conclude that an allowed speed deviation of \num{20}\% is the best compromise between platooning effects and individual properties for all approaches.
All approaches benefit from denser traffic producing more platooning opportunities.
The limited local knowledge of the \distributed{} approach does not depict an issue anymore with medium densities.

%

\section{Conclusion}%
\label{sec:conclusion}

\noindent
In this paper, we introduced and explored three approaches for \vtp{} assignments: \optimal{}, \centralized{}, and \distributed{}, using a \ac{MIP} solver and greedy heuristics, respectively.
Conceptually, the approaches differ in both knowledge about other vehicles as well as the used methodology.
We first define a similarity goal and, vice versa, the deviation among vehicles, thereby considering their individual requirements.
The aim is to increase vehicles' similarity, thus, minimizing their deviation in desired driving speed and position.
%
Our results show that all presented platooning approaches help in comparison to \human{} and \acc{} driving.
However, the selection of the formation algorithm is important, as it significantly influences the platoon assignment.
Our simulations show that the \optimal{} approach performs best in terms of individual platooning benefits but is conservative in terms of deviating from individual objectives.
In contrast, both presented greedy heuristics are less conservative and lead to a larger deviation from individual objectives.
We also see that the willingness to compromise can pay off as more vehicles can benefit from platooning.
While the \distributed{} approach has some disadvantages due to the limited local knowledge, it performs as good as the \optimal{} approach in most metrics.
Both outperform the \centralized{} approach, which suffers from synchronization and greedy selection effects.
The \optimal{} approach however assumes global knowledge and requires a complex \ac{MIP} solver to compute \vtp{} assignments.
Overall, the \distributed{} approach achieves close to optimal results but requires the least assumptions and complexity.
Therefore, we consider the \distributed{} approach the best approach among all presented approaches.

In future work, we want to investigate how a decentralized approach can increase vehicle knowledge while reducing synchronization effects at the same time.
Also, we plan to consider the heterogeneity of vehicles by using more of their properties such as their destination and vehicle capabilities within the similarity function.
Lastly, we want to consider more sophisticated maneuvers that allow merging of platoons and algorithms that allow to react to changes in properties by re-assigning vehicles.

%

\section{Acknowledgements}%
\label{sec:acks}

\noindent
The authors would like to thank Max Schettler, Dominik S. Buse, Agon Memedi, and Do\u{g}analp Ergen\c{c} for their support and feedback during the creation of this work.

%

\printbibliography{}

%

\begin{IEEEbiography}[{\includegraphics[width=1in,height=1.25in,clip,keepaspectratio]{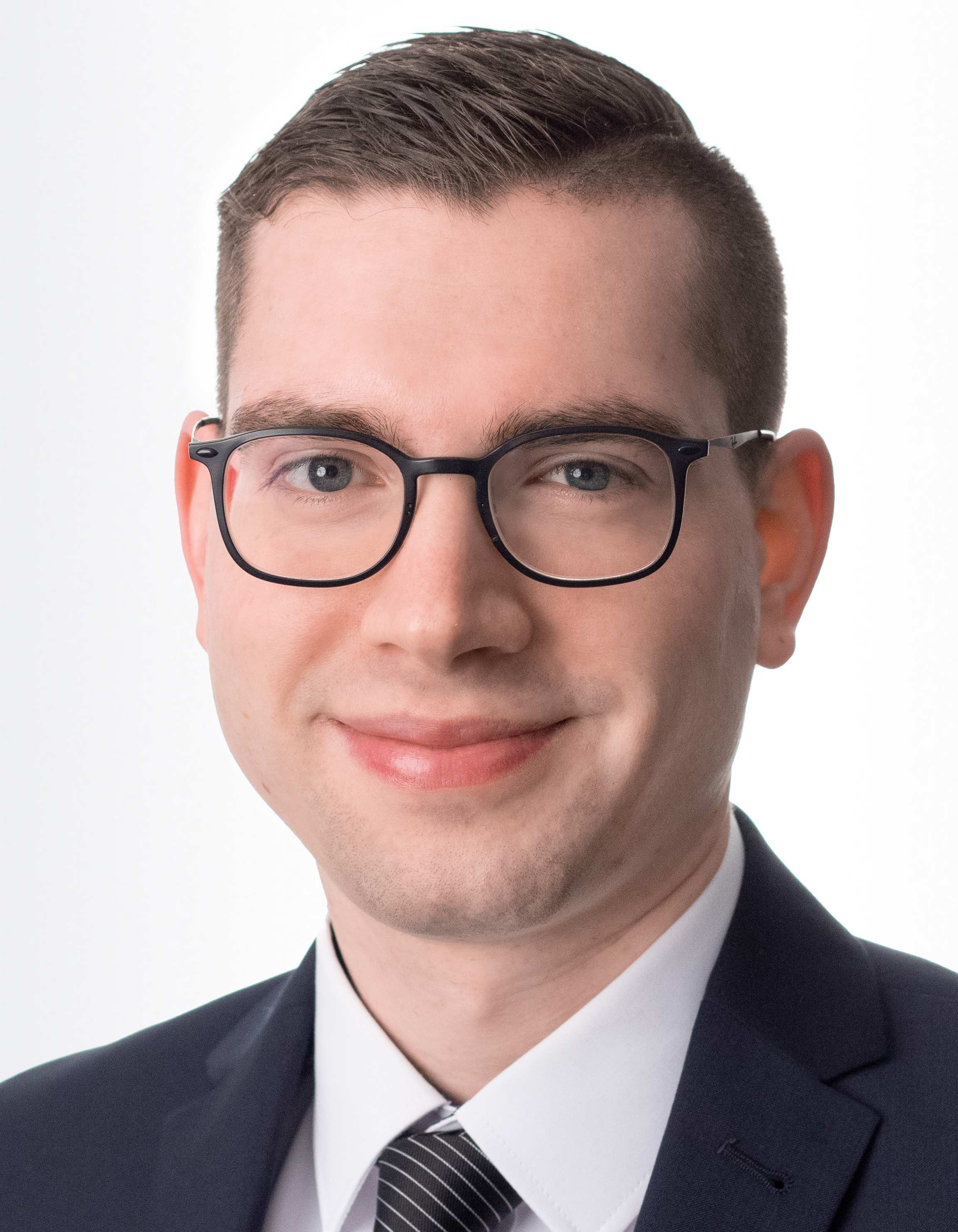}}]{Julian Heinovski}
(heinovski@ccs-labs.org)
is a PhD candidate and researcher at the Telecommunications Networks (TKN) group at the School of Electrical Engineering and Computer Science, TU Berlin, Germany.
He received his B.Sc.\ and M.Sc.\ degrees from the Dept.\ of Computer Science, Paderborn University, Germany, in 2016 and 2018, respectively.
Julian is an IEEE Graduate Student Member and an ACM Student Member as well as a Member of IEEE Intelligent Transportation Systems Society (ITSS) and IEEE Vehicular Technology Society (VTS).
He served as a reviewer for various manuscripts in the field of vehicular networks, cooperative driving, and intelligent transportation systems.
His research interest is in cooperative driving and intelligent transportation systems with a focus on vehicular platooning.
\end{IEEEbiography}

\begin{IEEEbiography}[{\includegraphics[width=1in,height=1.25in,clip,keepaspectratio]{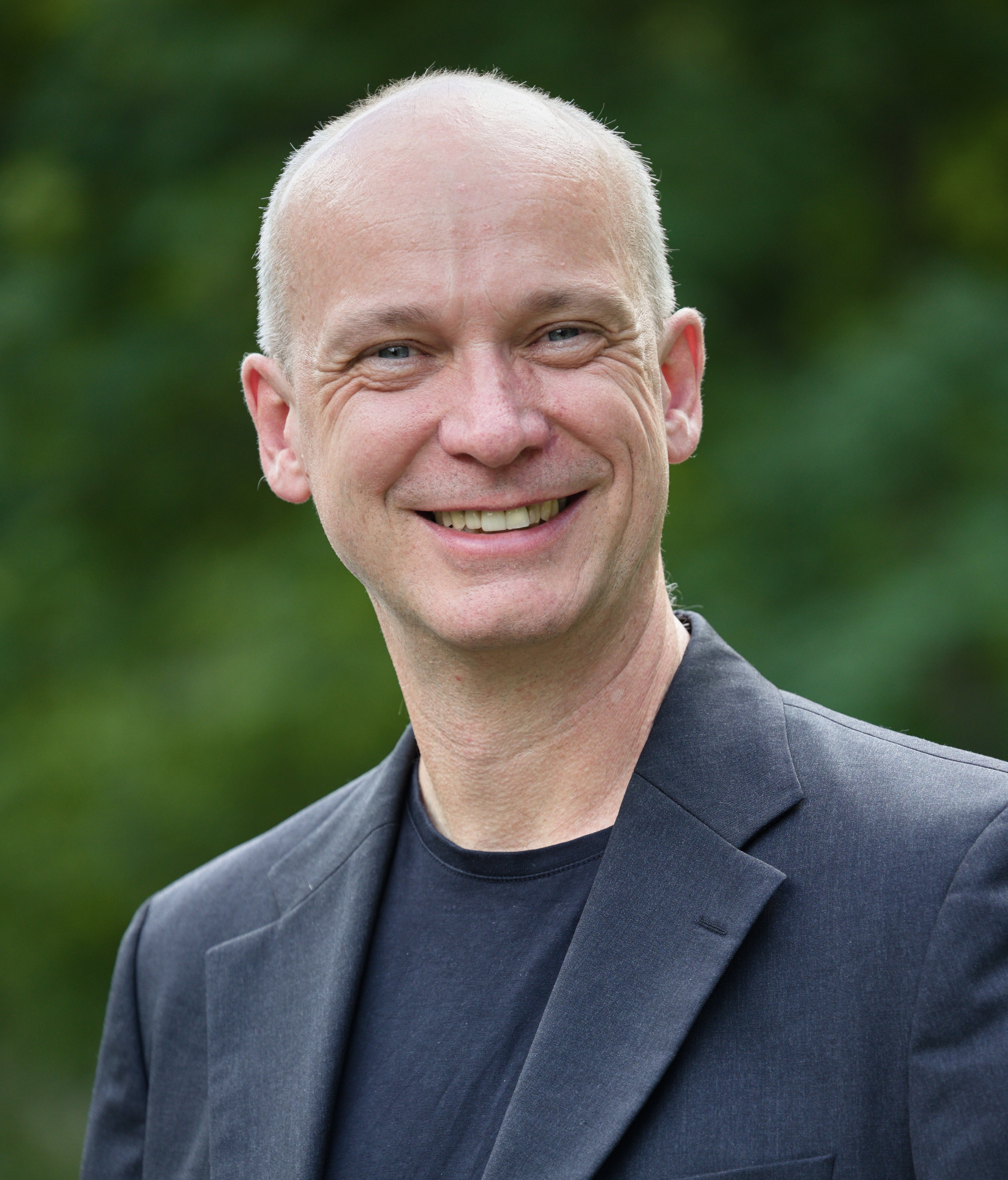}}]{Falko Dressler}
(dressler@ccs-labs.org)
is full professor and Chair for Telecommunication Networks at the School of Electrical Engineering and Computer Science, TU Berlin. He received his M.Sc. and Ph.D. degrees from the Dept. of Computer Science, University of Erlangen in 1998 and 2003, respectively.
Dr. Dressler has been associate editor-in-chief for IEEE Trans. on Mobile Computing and Elsevier Computer Communications as well as an editor for journals such as IEEE/ACM Trans. on Networking, IEEE Trans. on Network Science and Engineering, Elsevier Ad Hoc Networks, and Elsevier Nano Communication Networks. He has been chairing conferences such as IEEE INFOCOM, ACM MobiSys, ACM MobiHoc, IEEE VNC, IEEE GLOBECOM. He authored the textbooks Self-Organization in Sensor and Actor Networks published by Wiley \& Sons and Vehicular Networking published by Cambridge University Press. He has been an IEEE Distinguished Lecturer as well as an ACM Distinguished Speaker.
Dr. Dressler is an IEEE Fellow as well as an ACM Distinguished Member. He is a member of the German National Academy of Science and Engineering (acatech). He has been serving on the IEEE COMSOC Conference Council and the ACM SIGMOBILE Executive Committee. His research objectives include adaptive wireless networking (sub-6GHz, mmWave, visible light, molecular communication) and wireless-based sensing with applications in ad hoc and sensor networks, the Internet of Things, and Cyber-Physical Systems.
\end{IEEEbiography}

\end{document}